\documentclass[a4paper,fleqn,usenatbib,useAMS]{mnras}

\usepackage{graphicx}	% Including figure files
\usepackage{amsmath}	% Advanced maths commands
\usepackage{amssymb}	% Extra maths symbols
\usepackage{multicol}   % Multi-column entries in tables
\usepackage{bm}		% Bold maths symbols, including upright Greek
\usepackage{pdflscape}	% Landscape pages
\usepackage[export]{adjustbox}
\usepackage{cancel}
\newcommand{\ct}{\citealt}
\renewcommand{\thesubsubsection}{\thesubsection.\alph{subsubsection}}
\usepackage[T1]{fontenc}
\usepackage{ae,aecompl}

\usepackage{newtxtext,newtxmath}
\title[Hall Effect]{Hall Effect in Protostellar Disc Formation and Evolution}

\author[B. Zhao et al.]{Bo Zhao$^{1}$\thanks{Contact e-mail: \href{mailto:bo.zhao@mpe.mpg.de}{bo.zhao@mpe.mpg.de}},%\thanks{Present address: Giessenbachstr. 1, D-85748, Garching, Germany},
Paola Caselli$^{1}$,
Zhi-Yun Li$^{2}$,
Ruben Krasnopolsky$^{3}$,
Hsien Shang$^{3}$,
Ka Ho Lam$^{2}$\\
\\
$^{1}$Max-Planck-Institut f\"ur extraterrestrische Physik (MPE), Giessenbachstr. 1, 85748, Garching, Germany\\
$^{2}$University of Virginia, Astronomy Department, Charlottesville, USA, 22904\\
$^{3}$Academia Sinica Institute of Astronomy and Astrophysics (ASIAA), 10167, Taipei, Taiwan\\}
\pubyear{2019}

\begin{document}
\label{firstpage}
\pagerange{\pageref{firstpage}--\pageref{lastpage}}
\maketitle

\begin{abstract}
{The Hall effect is recently shown to be efficient in magnetized dense molecular 
cores, and could lead to a bimodal formation of rotationally supported discs 
(RSDs) in the first core phase. However, how such Hall dominated systems evolve 
in the protostellar accretion phase remains unclear. We carry out 2D 
axisymmetric simulations including Hall effect and Ohmic dissipation, 
with realistic magnetic diffusivities computed from our equilibrium chemical network. 
We find that %similar to ambipolar diffusion, 
Hall effect only becomes efficient when the large population of very small grains 
(VSGs: $\lesssim$100~$\AA$) is removed from the standard MRN size distribution. 
With such an enhanced Hall effect, however, the bimodality of disc formation does 
not continue into the main accretion phase. The outer part of the initial 
$\sim$40~AU disc formed in the anti-aligned configuration ($\bmath{\Omega \cdot B}<0$) 
flattens into a thin rotationally supported Hall current sheet as Hall effect 
moves the poloidal magnetic field radially inward relative to matter, 
leaving only the inner $\lesssim$10--20~AU RSD. In the aligned configuration 
($\bmath{\Omega \cdot B}>0$), disc formation is suppressed initially but a 
counter-rotating disc forms subsequently due to efficient azimuthal Hall drift. 
The counter-rotating disc first grows to $\sim$30~AU as Hall effect moves the 
magnetic field radially outward, but only the inner $\lesssim$10~AU RSD is 
long-lived like in the anti-aligned case. Besides removing VSGs, cosmic 
ray ionization rate should be below a few 10$^{-16}$~s$^{-1}$ for Hall effect 
to be efficient in disc formation. We conclude that Hall effect 
produces small $\lesssim$10--20~AU discs regardless of the polarity of the magnetic 
field, and that radially outward diffusion of magnetic fields remains crucial for 
disc formation and growth.}
\end{abstract}
%and magnetic braking remains efficient even with Hall effect reducing the level 
%of azimuthal bending of magnetic fields. 

\begin{keywords}
magnetic fields -MHD- circumstellar matter - stars: formation
\end{keywords}

%%%%%%%%%%%%%%%%%%%%%%%%%%%%%%%%%%%%%%%%%%%%%%%%%%

%%%%%%%%%%%%%%%%% BODY OF PAPER %%%%%%%%%%%%%%%%%%

\section{Introduction}
\label{Chap.Intro}

The formation of rotationally supported discs (RSDs) from magnetized dense 
molecular cloud cores, despite recent theoretical progress, remains a topic of 
debate in star formation. Tension still exists in how to resolve the so-called 
magnetic braking ``catastrophe'' \citep{Allen+2003b,MellonLi2008,HennebelleFromang2008}, 
which efficiently transports angular momentum away from the circumstellar region 
and suppresses disc formation in the axisymmetric ideal MHD limit. On the other hand, 
there has been ample observational evidence recently of Keplerian discs around young stellar 
objects \citep{WilliamsCieza2011,Tobin+2012,Tobin+2013,Segura-Cox+2016,Segura-Cox+2018}, 
which implies mechanisms that weaken magnetic braking should operate efficiently 
in collapsing cloud cores. 

Among different mechanisms proposed previously (for example, misalignment between 
the initial magnetic field and rotation axis, e.g., \ct{Joos+2012}; turbulence, 
e.g., \ct{Santos-Lima+2012}.), non-ideal MHD effects, especially ambipolar diffusion 
\citep[AD;][]{Masson+2015,Tomida+2015,Zhao+2016,Zhao+2018a} and Hall effect 
\citep{Tsukamoto+2015b,Wurster+2016}, are recently recognized as 
the most efficient and natural ways of averting the magnetic braking ``catastrophe'' 
and promoting disc formation. In fact, because dense cores are only slightly 
ionized \citep{Caselli+1998,BerginTafalla2007}, the flux freezing condition in the 
ideal MHD limit is not strictly satisfied, and decoupling of magnetic fields from the 
bulk fluid motion should commonly occur at various scales in collapsing cores. 
AD can be efficient in both the inner and outer envelopes \citep{Masson+2015,Zhao+2018a}, 
and Hall effect mainly dominates in the inner envelope 
\citep[$\sim$100~AU scale;][]{Tsukamoto+2015b,Tsukamoto+2017}. Since the pioneering 
work of \citet{Krasnopolsky+2011} and \citet{BraidingWardle2012a,BraidingWardle2012b}, 
Hall effect has recently been recognized by different groups as the dominant 
mechanism for disc formation \citep[e.g.,][]{Tsukamoto2016,WursterLi2018}. 
Particularly, such work often concludes that disc formation by Hall effect is bimodal; 
i.e., depending on the polarity of the magnetic field ($\bmath{B}$) with respect 
to the angular velocity vector ($\bmath{\Omega}$), disc formation can be suppressed when 
$\bmath{\Omega \cdot B}>0$ (aligned configuration) or promoted when 
$\bmath{\Omega \cdot B}<0$ (anti-aligned configuration). 

One common limitation of the existing work on Hall dominated core collapse 
is that the simulation usually stops shortly ($\lesssim$1~kyr) after the 
formation of the first hydrostatic core \citep{Larson1969}, as the Hall time 
step becomes intolerably small and/or Hall solver becomes highly unstable 
\citep[e.g.,][]{Tsukamoto+2015b,Tsukamoto+2017,Wurster+2018}. How does 
the disc-envelope system evolve subsequently in the main accretion phase 
remains unclear. \citet{Wurster+2016} in fact evolve the system somewhat 
longer in time ($\sim$5~kyr after the first core), and find the disc 
formed by Hall effect in the anti-aligned configuration appears to 
have a maximum radius at their intermediate time frame; however, they 
did not elaborate on this nor follow the system even further in time. 
Later, \citet{WursterBate2019} switch to a different code (without super 
time-stepping for Hall solver) and follow the long-term evolution of 
disc evolution and fragmentation, including both AD and Hall effect; 
yet Hall effect is not investigated alone.

More recently, \citet{Koga+2019} develop an analytical model for non-rotating cores, 
and predict growth of disc radius to $\sim$100~AU during the main accretion phase. 
However, as we will demonstrate below, the azimuthal Hall drift velocity of 
magnetic fields does not directly convert to the same amount of gas rotation. 
Instead, one should carefully consider the degree of azimuthal (instead of radial) 
field bending and the inward advection of poloidal magnetic fields, in order to 
estimate the magnetic torque and the rate of change of angular momentum. 
Nevertheless, their result of the existence of an optimal grain size for 
Hall diffusivity is consistent with what we found in \citet{Zhao+2018b}. 

Finally, different groups \citep{Dzyurkevich+2017,Zhao+2018b,Koga+2019} 
have confirmed the strong dependence of Hall diffusivity on microphysics, 
especially on the grain size distribution and cosmic-ray ionization rate.
Most of the recent work that show strong Hall effect in collapse simulations 
adopt $\sim$0.1~$\mu$m grains for the computation of Hall diffusivity 
\citep[e.g.,][]{Tsukamoto+2015b,Wurster+2016,Wurster+2018}. In comparison, 
\citep{Li+2011} have tested the MRN \citep[Mathis-Rumpl-Nordsieck;][]{Mathis+1977} 
size distribution and singly-sized 1~$\mu$m grain, and instead find that Hall 
effect is inefficient in affecting disc formation. As we will show in this paper, 
neither the MRN distribution nor the 1~$\mu$m grain is favourable for producing a 
large Hall diffusivity; a slightly evolved grain size distribution free of very 
small grains (VSGs: $\lesssim$100~$\AA$) can greatly promote disc formation by Hall 
effect. 

Considering that Hall effect is numerically demanding, we conduct a 
parameter study of the Hall effect on disc formation and evolution 
using two-dimensional (2D) axisymmetric simulations with realistic 
Hall diffusivity computed from our chemical network. We explored different 
magnetic field strengths, rotation speeds, grain sizes, and cosmic-ray 
ionization rates. For the first time, we show that the bimodality of disc 
formation by Hall effect no longer holds into the main accretion phase, and 
that the eventual RSD radius is smaller than 20~AU, surrounding which is 
a flattened Hall current sheet. The rest of the paper is organized as follows. 
Section \ref{Chap.Hall} demonstrates the basic principles of Hall effect 
and the Hall drift in both radial and azimuthal directions. 
Section \ref{Chap.IC} describes the initial conditions of the simulation set, 
together with an overview of the results. In Section \ref{Chap.SimulResult}, 
we present and analyze the simulation results, emphasizing the microscopic 
conditions for efficient Hall effect and the subsequent evolution of discs 
formed by Hall effect. We discuss possible implications for protoplanetary discs 
and connect to recent observations in Section \ref{Chap.Discuss}. Finally, 
we summarize the results in Section \ref{Chap.Summary} and leave further 
study including both AD and Hall effect in a consecutive paper (Paper II hereafter).

\section{Hall Effect}
\label{Chap.Hall}

Considering only the Hall Effect and Ohmic dissipation, the evolution of the magnetic 
field $\bmath{B}$ is governed by the following magnetic induction equation, 
\begin{equation}
\label{Eq:induct}
\begin{split}
{\partial \bmath{B} \over \partial t} & = \nabla \times (\bmath{\varv} \times \bmath{B}) - \nabla \times \left\{\eta_{\rm H}(\nabla \times \bmath{B}) \times {\bmath{B} \over B} + \eta_{\rm O}\nabla \times \bmath{B}\right\}\\
& = \nabla \times \left[(\bmath{\varv} + \bmath{\varv}_{\rm H}) \times \bmath{B} - \eta_{\rm O}\nabla \times \bmath{B}\right]~,
\end{split}
\end{equation}
where $\bmath{\varv}$ is the fluid velocity, $\eta_{\rm O}$, $\eta_{\rm H}$ 
are the Hall and Ohmic diffusivities, respectively, and $\bmath{\varv}_{\rm H}$ 
denotes the drift velocity of the magnetic field induced by Hall effect, 
which is defined as, 
\begin{equation}
\label{Eq:v_H}
\bmath{\varv}_{\rm H} = -\eta_{\rm H} {\nabla \times \bmath{B} \over B} = -\eta_{\rm H} {4\pi \bmath{J} \over c B}~,
\end{equation}
where $c$ is the light speed, and $\bmath{J}$ is the electric current. 
We also denote $\bmath{\varv}_{\rm iH}=\bmath{\varv}+\bmath{\varv}_{\rm H}$ as 
the velocity of the charged species that dominates the Hall diffusivity 
(e.g., electrons in an ion-electron dominated plasma). 
%that is well coupled to %the magnetic field, i.e., $\bmath{\varv}_{\rm iH}$ 
%represents the motion of the magnetic field lines dragged by the dominant 
%charged species due to bulk fluid motion and Hall drift.
One can further define an effective velocity of the magnetic 
field lines, taking Ohmic dissipation into account,\footnote{Note that Ohmic 
dissipation, dominating the diffusion of magnetic fields at very high densities 
($\gtrsim$10$^{13}$~cm$^{-3}$), does not directly drift the magnetic field lines, 
but instead modifies the electromotive force (EMF) by dissipating the total electric 
current.} as,
\begin{equation}
\bmath{\varv}_{\rm B} = {c \over B^2} \bmath{E \times B} = \bmath{\varv}_{\rm iH,\perp} + {\eta_{\rm O} \over B^2} (\nabla \times \bmath{B}) \times \bmath{B}~
\end{equation}
\citep[e.g.,][]{KunzMouschovias2009,BraidingWardle2012a}, in which $\bmath{E}$ 
is the electric field and $\bmath{\varv}_{\rm iH,\perp}$ represents the component of 
$\bmath{\varv}_{\rm iH}$ perpendicular to the magnetic field lines.

We first discuss the sign of the Hall diffusivity $\eta_{\rm H}$. 
%and the direction of the Hall drift velocity $\bmath{\varv}_{\rm H}$. 
In an ion-electron dominated plasma, $\eta_{\rm H} = {c B \over 4\pi e n_{\rm e}}$ 
and $\bmath{\varv}_{\rm H} = - {\bmath{J} \over e n_{\rm e}}~$, meaning 
that the Hall drift of the magnetic field is equivalent to the mean 
velocity of the electrons associated with the current they carry 
relative to ions that are assumed to move together with the neutrals. 
However, this does not always hold in a weakly-ionized medium with 
charged sub-micron grains \citep{Dzyurkevich+2017,Zhao+2018b,Koga+2019}.

More generally, the Hall diffusivity $\eta_{\rm H}$ is related to the components 
of fluid conductivities as:
\begin{equation}
\eta_{\rm H} = {c^2 \over 4\pi} {\sigma_{\rm H} \over \sigma_{\rm H}^2 + \sigma_{\rm P}^2}~,
\end{equation}
where $\sigma_{\rm P}$ (Pederson conductivity) and $\sigma_{\rm H}$ (Hall conductivity) 
are components of the conductivity tensor, given by:
\begin{equation}
\label{Eq:sgmP}
\sigma_{\rm P} = {e c \over B} \sum_i {{Z_i n_i \beta_{i, \rm H_2}} \over {1+\beta_{i,\rm H_2}^2}}~,
\end{equation}
\begin{equation}
\label{Eq:sgmH}
\sigma_{\rm H} = -{e c \over B} \sum_i {{Z_i n_i \beta_{i,\rm H_2}^2} \over {1+\beta_{i,\rm H_2}^2}}~,
\end{equation}
\citep{NormanHeyvaerts1985, WardleNg1999}, in which $n_i$ is the number density 
of charged species $i$, and $\beta_{i,\rm H_2}$ is the usual Hall parameter 
for species $i$ that characterizes the relative magnitude of 
Lorentz and drag forces (e.g., \ct{Wardle2007}; \ct{Zhao+2016}). 
Note that we use a slightly different formulation of the Hall conductivity 
$\sigma_{\rm H}$ in Eq.~\ref{Eq:sgmH} than \citet{Zhao+2016,Zhao+2018b}; 
but they are essentially the same and can be validated using the charge 
neutrality condition $\Sigma Z_i n_i = 0$ \citep{Wardle2007}. 
In fact, the form of $\sigma_{\rm H}$ in Eq.~\ref{Eq:sgmH} has a more consistent 
physical meaning with the one in an ion-electron dominated plasma, in that 
$\sigma_{\rm H} > 0$ (and hence $\eta_{\rm H} > 0$) when electron dominates 
the Hall conductivity. 

We have shown in \citet{Zhao+2018b} that Hall diffusivity is mostly negative 
in the low density envelope \citep[see also][]{Nakano+2002,Tsukamoto+2017,Dzyurkevich+2017}; 
physically, $\eta_{\rm H}$ in such density regimes is mainly dominated by 
positively charged species instead of electrons.\footnote{The interpretation 
here is also different from the analysis in \citet{Zhao+2018b} because of the 
different formulations of $\sigma_{\rm H}$.} The sign of $\eta_{\rm H}$ usually 
changes to positive at very high densities ($\gtrsim$10$^{13}$~cm$^{-3}$), 
where even ions can become decoupled from the magnetic field 
\citep{KunzMouschovias2010} and electrons dominate the Hall conductivity.

\subsection{Hall Drift of Magnetic Fields}
\label{S.HallDrift}

\begin{figure*}
\includegraphics[width=\textwidth]{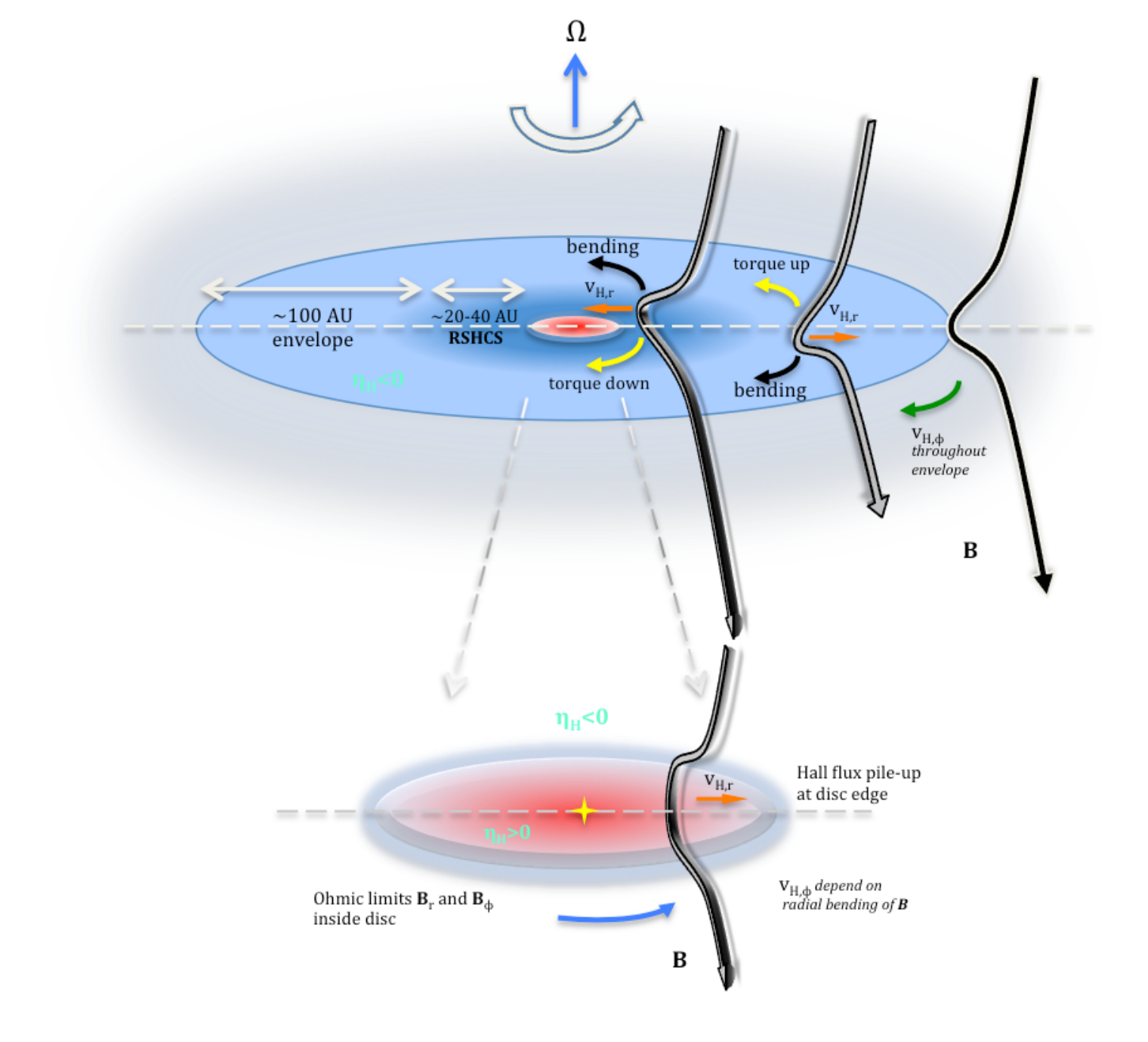}
\caption{Sketch of magnetic field morphologies in the inner envelope (top) 
and in the disc (bottom) for the anti-aligned configuration 
($\bmath{\Omega}\cdot\bmath{B}<0$) with Hall effect and Ohmic dissipation. 
$\varv_{{\rm H},\phi}$ and $\varv_{{\rm H},r}$ are the azimuthal and radial Hall 
drift velocities, respectively; $\eta_{\rm H}$ denotes the Hall diffusivity. 
The structure outside the disc is termed as RSHCS (``rotationally supported 
Hall current sheet'') and will be discussed in \S~\ref{S.HallSheet}.}
\label{Fig:sketch-B}
\end{figure*}

\begin{figure*}
\includegraphics[width=\textwidth]{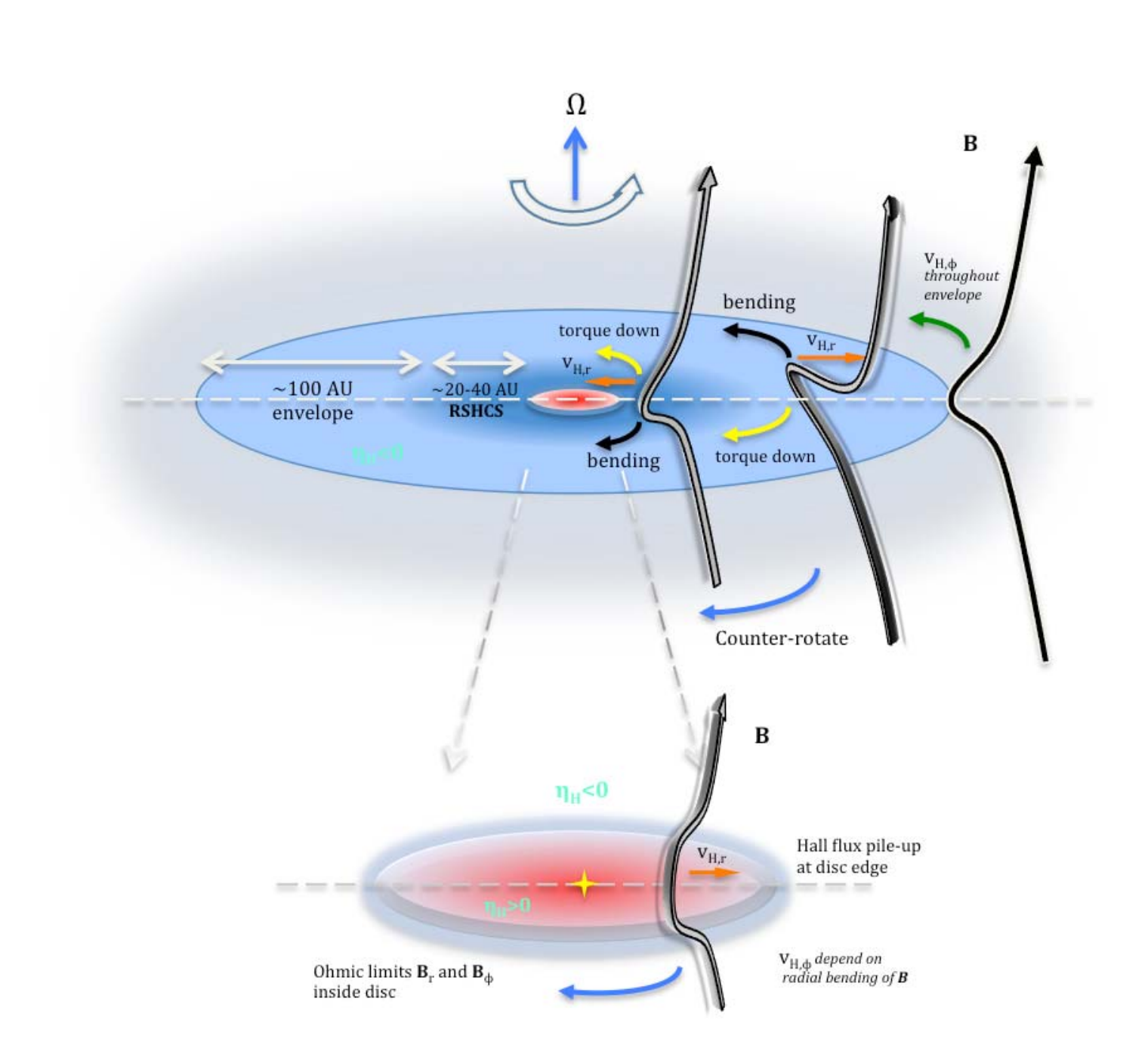}
\caption{Same as Fig.~\ref{Fig:sketch-B}, but for the aligned configuration 
($\bmath{\Omega}\cdot\bmath{B}>0$).}
\label{Fig:sketch+B}
\end{figure*}

In the usual picture of disc formation induced by Hall effect at the 
envelope scale, the poloidal magnetic fields that are the most curved across 
the pseudo-disc (normally the equatorial plane) produces an azimuthal current 
$J_{\phi}={c \over 4\pi}(\nabla \times \bmath{B})_\phi$, and the corresponding 
Hall drift along the azimuthal direction, in cylindrical coordinates ($r$,$\phi$,$z$), 
is:
\begin{equation}
\label{Eq:v_Hphi}
\varv_{{\rm H},\phi} = -\eta_{\rm H} {4\pi J_{\phi} \over c B} \approx -{\eta_{\rm H} \over B}{\partial B_r \over \partial z}~,
\end{equation}
in which we neglect the contribution from the pressure term 
$\partial B_z \over \partial r$ that is many orders of magnitude smaller 
\citep{Zhao+2018a}. As a result, magnetic fields drift with velocity 
$\bmath{\varv}_{{\rm H},\phi}$ that is along the direction of the azimuthal 
current $J_{\phi}$ (with $\eta_{\rm H}<0$). For example, in the anti-aligned 
configuration (Fig.~\ref{Fig:sketch-B}), the magnetic field is originally 
bended azimuthally by the gas rotation towards $-J_{\phi}$ in the envelope, 
producing a magnetic tension force along $J_{\phi}$. Thus, 
the azimuthal Hall drift will weaken such an azimuthal bending and the 
corresponding magnetic braking. In situations where the Hall drift is large 
enough to flip the direction of the azimuthal bending of magnetic fields 
(``over-bending'' of magnetic fields, from bending towards $-J_{\phi}$ to $J_{\phi}$; 
e.g., the middle field line of the upper panel of Fig.~\ref{Fig:sketch-B}), 
the magnetic tension force reverses direction and the magnetic braking instead becomes 
a magnetic spin-up of the original gas rotation along $-J_{\phi}$. On the other hand, 
if azimuthally the original field bending (gas rotation) is along the same direction 
as the Hall drift (towards $J_{\phi}$) such as in the aligned configuration 
(the middle field line of the upper panel of Fig.~\ref{Fig:sketch+B}), 
the magnetic tension force pointing $-J_{\phi}$ is enhanced by such a Hall drift, 
which can halt entirely the original gas rotation as well as further spin the gas 
in the opposite direction along $-J_{\phi}$ (``over-shooting'' of gas rotation: 
counter-rotation). Note that the azimuthal Hall drift is sensitive to the radial 
pinchness of magnetic field lines across the pseudo-disc, which will be weakened by 
the radial ambipolar drift (see Paper II for detailed discussions).

However, as toroidal magnetic fields develop in the inner envelope 
(10$^2$~AU scale) where rotation, either due to the original cloud 
rotation or the Hall spin-up, becomes much more important than infall, 
the toroidal magnetic fields pinched in the azimuthal direction across 
the pseudo-disc induces a radial current 
$J_{r}={c \over 4\pi}(\nabla \times \bmath{B})_r$, and accordingly the 
radial Hall drift is:
\begin{equation}
\label{Eq:v_Hr}
\varv_{{\rm H},r} = -\eta_{\rm H}{4\pi J_r \over c B} \approx {\eta_{\rm H} \over B}{\partial B_\phi \over \partial z}~,
\end{equation}
\citep[similar to][]{BaiStone2017}, in which the contribution from the magnetic 
pressure along the azimuthal direction $-{1 \over r}{\partial B_z \over \partial \phi}$ 
is negligible. In the inner envelope (mostly $\eta_{\rm H}<0$), magnetic fields 
drift with velocity $\bmath{\varv}_{{\rm H},r}$ that is along the direction of 
the radial current $J_{r}$. Particularly, the radial Hall drift 
$\bmath{\varv}_{{\rm H},r}$ is always along $+\bmath{r}$ (radially outward) 
if the azimuthal bending of magnetic fields goes the same direction as the 
above-mentioned azimuthal Hall drift $\bmath{\varv}_{{\rm H},\phi}$ (e.g., 
``over-bending'' in the anti-aligned configuration), regardless of the polarity 
of the poloidal magnetic field and the sign of $\eta_{\rm H}$ (squared for both 
$\eta_{\rm H}$ and $\bmath{B}$). In this case, the radial Hall drift tends 
to reduce the radial pinchness of the poloidal magnetic field lines, and thus 
places a negative feedback to the Hall drift in the azimuthal direction.

It is worth pointing out that, the case discussed above mainly applies to 
collapse dominated regions where magnetic fields are primarily bended 
radially, so that azimuthal drift is the dominant Hall drift mode. 
Otherwise, in regions where magnetic fields are mostly bended azimuthally 
\citep[e.g., protoplanetary discs,][]{BaiStone2017}, radial Hall drift will 
be the dominant mode, whereas azimuthal Hall drift becomes a secondary effect 
and generally (along with AD) provides negative feedback to the radial Hall drift.

\subsection{Magnetic Braking Torque}
\label{S.Torque}

Before presenting the result, we briefly investigate the magnetic torque 
caused by bending magnetic field lines, which will be the key to understand 
the magnetic braking and disc evolution. The magnetic torque $\bmath{N}_m$ 
for any material at a given radius $r$ is,
\begin{equation}
\label{Eq:Nm}
\bmath{N}_{\rm m} = \bmath{r} \times \bmath{F}_{\rm L} = \bmath{r} \times \left[{(\nabla \times \bmath{B}) \times \bmath{B} \over 4\pi}\right]~,
\end{equation}
in which $\bmath{F}_{\rm L}$ is the Lorentz force. 
Along the pseudo-disc plane, the $z$-component of the magnetic torque that is 
responsible for the change of angular momentum $\bmath{L}$ along z-direction 
can be expressed, keeping only the leading terms, as,
\begin{equation}
\label{Eq:Lz}
{{\rm d}L_z \over {\rm d}t} = N_{{\rm m},z} = r F_{L,\phi} \approx {1 \over 4\pi}\left(B_r B_\phi + r B_z{\partial B_\phi \over \partial z}\right)~.
\end{equation}
In general, the second term ($\propto r B_z J_r$) is much larger than the 
first term, since $B_z$ is usually the dominant component of magnetic field 
in our set-up. Therefore, for magnetic braking to be inefficient, both the 
poloidal magnetic field ($B_z$) and the azimuthal bending of the magnetic 
field ($\propto J_r$) have to be small. Qualitatively, AD radially reduces 
the total magnetic flux arrived in the inner envelope \citep{Zhao+2018a}, 
hence reducing $B_z$ therein; Hall drift mainly regulates the azimuthal 
bending of the magnetic field, affecting $J_r$ \citep{Tsukamoto+2017,Wurster+2018}. 
Usually, the radial ambipolar and azimuthal Hall drift together affects 
the magnetic braking efficiency. However, the radial Hall drift can dominate 
the radial drift of magnetic fields once the azimuthal bending of magnetic 
fields becomes severe. 
Therefore, to fully understand the behavior of magnetic field and the 
efficiency of magnetic braking, different components of both ambipolar and 
Hall drift have to be considered, which will be discussed in more details 
in Paper II. It is worth noting that the bending direction of magnetic 
field lines is not necessarily the same as the direction of drift velocities 
(displacement versus velocity), for both AD and Hall effect, along either 
radial or azimuthal direction.

From Eq.~\ref{Eq:Lz}, we can also estimate the change of azimuthal velocity 
$\varv_\phi$ due to the magnetic force as,
\begin{equation}
\label{Eq:dv_phi}
{{\rm d} \varv_\phi \over {\rm d} t} \approx {1 \over 4\pi\rho}\left({B_r B_\phi \over r}+ B_z{\partial B_\phi \over \partial z}\right) \approx {1 \over 4\pi\rho}B_z{\partial B_\phi \over \partial z}~.
\end{equation}
in which the radial advection of specific angular momentum 
${\varv_r \over r}{\partial (r\varv_\phi) \over \partial r}$ 
is hidden in the convective derivative. If we ignore the pressure gradient 
and gravitational force in the azimuthal direction, Eq.~\ref{Eq:dv_phi} 
shows that the time evolution of gas rotation velocity is determined by 
the magnetic torque (i.e., the poloidal magnetic field strength and the 
azimuthal bending of magnetic field lines), and that $\varv_\phi$ may not be 
directly approximated by the azimuthal Hall drift velocity \citep{Koga+2019} 
that mainly depends on the radial pinch of the magnetic field (Eq.~\ref{Eq:v_Hphi}). 
Instead, the radial Hall drift velocity $\varv_{{\rm H}, r}$ 
($\propto {\partial B_\phi \over \partial z}$) is a more relevant quantity, 
in that ${{\rm d} \varv_\phi \over {\rm d} t} \approx {1 \over 4\pi\rho} {B_z B \over \eta_{\rm H}} \varv_{{\rm H}, r}$.

\section{Initial Condition}
\label{Chap.IC}

To investigate the Hall effect in disc formation and subsequent evolution, 
we carry out two-dimensional (2D) numerical simulations using 
ZeusTW code \citep{Krasnopolsky+2010}, which adopts explicit method 
\citep{SanoStone2002} and sub-cycling \citep{Huba2003} to treat the Hall term 
in the induction equation. The magnetic diffusivities are obtained by linearly 
interpolating the tabulated chemical network of \citet{Zhao+2018b} with molecular 
freeze-out/desorption process turned on, which slightly enhances Hall 
effect at number densities between $\sim$10$^{9}$--10$^{11}$~cm$^{-3}$ than 
that without freeze-out \citep[due to slightly fewer charged ions and grains;][]{Zhao+2018b}.
%which guarantees the condition $\eta_{\rm H}<0$ in the infalling envelope 
%for the grain size distributions adopted in this study.

The initial conditions are similar to the set-ups in \citet{Zhao+2018a}. 
We initialize a uniform, isolated spherical core with total mass 
$M_{\rm c}=1.0~M_{\sun}$, and radius $R_{\rm c}=10^{17}$~cm~$\approx 6684$~AU. 
This corresponds to an initial mass density 
$\rho_0=4.77 \times 10^{-19}$~g~cm$^{-3}$ and a number density for 
molecular hydrogen $n({\rm H}_2)=1.2 \times 10^5$~cm$^{-3}$ 
(assuming mean molecular weight $\mu=2.36$). The free-fall time of the core 
is about $t_{\rm ff} = 3 \times 10^{12}$~s~$\approx 9.6 \times 10^4$~yr. 
We adopt the same barotropic equation of state as that in \citet{Zhao+2018a}, 
with a smooth transition from isothermal to adiabatic phase to mimic the 
compressional heating of the gas \citep{Tomida+2013}. 
The core is rotating initially as a solid-body with angular speed 
$\omega_0=1 \times 10^{-13}$~s$^{-1}$ for slow rotating case, and 
$2 \times 10^{-13}$~s$^{-1}$ for fast rotating case, which corresponds to 
a ratio of rotational to gravitational energy $\beta_{\rm rot}\approx$ 0.025 and 0.1,
respectively (consistent with the typical $\beta_{\rm rot}$ values from 
\ct{Goodman+1993} and \ct{Caselli+2002b}). 

Because the Hall effect depends on the direction of the magnetic field, 
we consider both the aligned ($\bmath{\Omega}\cdot\bmath{B}>0$) and 
anti-aligned ($\bmath{\Omega}\cdot\bmath{B}<0$) configuration 
between magnetic field and angular momentum directions, which have 
benn shown to suppress or promote disc formation in the early phases 
\citep[e.g.,][]{Tsukamoto+2015b,Wurster+2016}. 
The magnetic field strength $B_0$ is of 42.5~$\mu$G for strong field case 
and 21.3~$\mu$G for weak field case, which gives a dimensionless mass-to-flux ratio 
$\lambda$ ($\equiv {M_{\rm c} \over \pi R_{\rm c}^2 B_0}2\pi\sqrt{G}$) of 2.4 and 4.8, 
respectively. It is consistent with the mean value of $\lambda$$\sim$2 inferred 
in cloud cores with densities $\sim$10$^3$--10$^4$~cm$^{-3}$ 
from the OH Zeeman observations by \citet{TrolandCrutcher2008}.

We adopt the spherical coordinate system ($r$, $\theta$, $\phi$) that better 
conserves angular momentum than Cartesian coordinate system in such a 
collapse problem. The grid is non-uniform, providing high resolution 
towards the innermost region of simulation domain. The inner boundary has 
a radius $r_{\rm in}=3 \times 10^{13}$~cm~$=2$~AU and the outer has 
$r_{\rm out}=10^{\rm 17}$~cm. At both boundaries, we 
impose a standard outflow boundary conditions to allow matter to leave 
the computational domain. The mass accreted across the inner boundary 
is collected at the centre as the stellar object. We use a total of 
$120 \times 96$ grid points. The grid is non-uniform in the 
$\theta$-direction with $\delta \theta=0.6713^{\circ}$ near the equator, 
and non-uniform in the $r$-direction with a spacing $\delta r=0.2$~AU 
next to the inner boundary. The $r$-direction spacing increases geometrically 
outward by a constant factor of $\sim$1.0663, and the $\theta$-direction 
spacing increases geometrically from the equator to either pole by a 
constant factor of $\sim$1.0387.

The fractional abundances of charged species are precomputed and tabulated 
using the chemical code presented in \citet{Zhao+2018b} for obtaining 
the magnetic diffusivities. We adopt a cosmic-ray (CR) ionization rate 
of $\zeta_0^{\rm H_2}=10^{-17}$~s$^{-1}$ (and $10^{-16}$~s$^{-1}$ for 
several comparison models) at the cloud edge with a characteristic 
attenuation length of $\sim$200~g~cm$^{-2}$
\citep[][by fitting their Fig. 8]{Padovani+2018}. 
We choose mainly four different grain size distributions of MRN-type, 
with fixed power law index 
of $-3.5$ and maximum grain size of $a_{\rm max}=0.25~\mu$m, but different 
minimum grain size $a_{\rm min}$=0.005 (MRN), 0.01 (min1), 0.03 (opt3), 
and 0.1~$\mu$m (trMRN), respectively. 
We also consider a large singly-sized grain case with $a=1.0~\mu$m (LG). 
As shown in \citet{Zhao+2018b}, there exists a favourable $a_{\rm min}$ 
between 0.03--0.04~$\mu$m for which Hall diffusivity reaches an 
optimal level in the inner envelope of the collapsing core; 
either increasing or decreasing $a_{\rm min}$ will suppress the Hall 
diffusivity $\eta_{\rm H}$ (either the Pederson conductivity becomes 
large or both the Hall and Pederson conductivity become large).

Due to the existence of short-wavelength whistler waves, explicit method 
for solving the Hall MHD equations can be generally unstable. Different 
numerical schemes have been developed to improve the stability of the Hall MHD, 
including higher order explicit method \citep[e.g.,][]{KunzLesur2013}, 
implicit method \citep{Falle2003,Toth+2008}, special operator-splitting 
\citep{OSullivanDownes2007,Bai2014}, and whistler modified Riemann solver 
\citep{Toth+2008,Lesur+2014,Marchand+2018}. The explicit method adopted 
in this study is second-order, the growth rate of the numerical instability 
can be reduced more substantially than in first-order method, as Hall time step 
decreases and/or with the introduction of additional physical or numerical 
diffusivities. %To ensure the numerical stability of the Hall solver, 
Therefore, we turn on the Ohmic dissipation with a varying floor 
equals to the smaller of 10$^{18}$~cm$^2$~s$^{-1}$ and $\eta_{\rm H}$. 
Such a resistivity floor is low enough to not noticeably weaken 
the electric current density as well as the Hall effect in the inner envelope 
(see discussions of \ct{Krasnopolsky+2011}). Furthermore, as will be discussed 
in \S~\ref{S.Ohmic}, the rapid increase of $\eta_{\rm O}$ in the disc will 
significantly limit both the radial ($B_r$) and azimuthal ($B_\phi$) components 
of the magnetic field, leading to strong suppression of Hall effect in the disc. 
In this regard, we impose a relatively small d$t$ floor for Hall effect 
with d$t_{\rm floor, H} \sim 5\times10^4$~s, which caps the Hall 
diffusivity\footnote{\label{foot:dt_H}The cap of $\eta_{\rm H}$ is computed for each 
cell as CFL${|\delta x|_{\rm min}^2 \over 4 {\rm d}t_{\rm floor}}$, where CFL 
is the Courant-Friedrichs-Lewy number that is set to 0.2, 
and $|\delta x|_{\rm min}$ is the smallest of the cell's sizes along 
$r$ and $\theta$ directions.} %yet only takes effect 
within $\lesssim$10~AU where Ohmic dissipation already dominates 
and in a small region of the bipolar cavity. 
Note that limiting $\eta_{\rm H}$ actually reduces the whistler wave speed 
$c_{\rm w}$ ($\propto {\eta_{\rm H} \over |\delta x|_{\rm min}}$; see Appendix~\ref{App.B}) 
and relaxes the requirement for small time steps to stablize the Hall MHD. 
Finally, we set a global ceiling of Hall diffusivity using $Q_0 = 3\times10^{22}$ 
in the Lorentz-Heaviside unit (maximum value used in \ct{Krasnopolsky+2011}; 
$\eta_{\rm H} = Q|B|/\sqrt{4\pi}$), which avoids unnecessary computation cost 
at 10$^3$~AU scale.

With different magnetic field strength (in terms of $\lambda$), grain 
size distribution (in terms of $a_{\rm min}$), and either aligned or anti-aligned 
magnetic field configuration, we summarize a total of 34 numerical simulations 
including 2 non-rotating models, 4 slow rotating models, and 4 high 
$\zeta^{\rm H_2}$ models in Table~\ref{Tab:model1} and \ref{Tab:model2}, 
and 2 models with very weak magnetic field strength $\lambda \sim 9.6$ in 
Table~\ref{Tab:model3}. Note that 2 models (2.4trMRN-AO and 4.8trMRN-AO) from 
\citet{Zhao+2016} considering only AD and Ohmic dissipation are also listed, 
to be used for analysis and comparison. 
\begin{table}
%\begin{minipage}{80mm}
\caption{Model Parameters for strong B-field $B_0\approx42.5~\mu$G ($\lambda$$\sim$2.4)}
\label{Tab:model1}
\resizebox{1.1\columnwidth}{!}{
\begin{tabular}{lcccc}
\hline\hline
Model$^\ddagger$ & Grain Size$^\dagger$ & $\zeta_0^{\rm H_2}$ & $\beta_{\rm rot}$ & Radius \& Morphology$^\ast$ \\
& Dist. & (10$^{-17}$~s$^{-1}$) & & (AU) \\
\hline
2.4MRN-H$^-$O & MRN & 1 & 0.1 & $\sim$8(FHSC)$\downarrow$<2 \\
2.4MRN-H$^+$O & MRN & 1 & 0.1 & <2 \\
\hline
2.4min1-H$^-$O & min1 & 1 & 0.1 & $\uparrow${\bf 40--50}$\downarrow$$\lesssim${\bf 16} \\
2.4min1-H$^+$O & min1 & 1 & 0.1 & $\downarrow$<2 $\Rightarrow$ {\bf 30--40}$^\circlearrowright$$\downarrow$$\lesssim${\bf 10}$^\circlearrowright$\\
\hline
2.4opt3-H$^-$O & opt3 & 1 & 0.1 & $\uparrow${\bf 40--50}$\downarrow$$\lesssim${\bf 16} \\
2.4opt3-H$^+$O & opt3 & 1 & 0.1 & $\downarrow$<2 $\Rightarrow$ {\bf 30--40}$^\circlearrowright$$\downarrow$$\lesssim${\bf 10}$^\circlearrowright$ \\
2.4opt3-H$^-$O-Slw & opt3 & 1 & 0.025 & $\uparrow${\bf 40--50}$\downarrow$$\lesssim${\bf 16} \\
2.4opt3-H$^+$O-Slw & opt3 & 1 & 0.025 & $\downarrow$<2 $\Rightarrow$ {\bf 30--40}$^\circlearrowright$$\downarrow$$\lesssim${\bf 12}$^\circlearrowright$ \\
2.4opt3-HO-NoRot & opt3 & 1 & 0 & $\uparrow${\bf 30--40}$^\circlearrowright$$\downarrow${\bf 16}$^\circlearrowright$ \\
\hline
2.4trMRN-AO & trMRN & 1 & 0.1 & {\bf 30--40} (Disc+Spiral/Ring) \\
2.4trMRN-H$^-$O & trMRN & 1 & 0.1 & $\uparrow$$\lesssim${\bf 40}$\downarrow$$\lesssim${\bf 10} \\
2.4trMRN-H$^+$O & trMRN & 1 & 0.1 & $\downarrow$<2 $\Rightarrow$ {\bf 30--40}$^\circlearrowright$$\downarrow$$\lesssim${\bf 10}$^\circlearrowright$ \\
\hline
2.4LG-H$^-$O & LG & 1 & 0.1 & <2 \\
2.4LG-H$^+$O & LG & 1 & 0.1 & $\sim$5(FHSC)$\downarrow$<2 \\
\hline
2.4CR10opt3-H$^-$O & opt3 & 10 & 0.1 & $\uparrow$$\lesssim${\bf 10}$\downarrow$$\lesssim$8 \\
2.4CR10opt3-H$^+$O & opt3 & 10 & 0.1 & <2 $\Rightarrow$ $\lesssim$7$^\circlearrowright$ \\
\hline\hline
\end{tabular}
}
\\
$\dagger$~MRN: full MRN distribution with $a_{\rm min}$=0.005~$\mu$m \\
$\dagger$~min1: truncated MRN distribution with $a_{\rm min}$=0.01~$\mu$m \\
$\dagger$~opt3: truncated MRN distribution with $a_{\rm min}$=0.03~$\mu$m, 
with which Hall diffusivity reaches an optimal level in the inner envelope \\
$\dagger$~trMRN: truncated MRN distribution with $a_{\rm min}$=0.1~$\mu$m \\
$\dagger$~LG: singly-sized grains with $a$=1.0~$\mu$m; note that LG models have $\eta_{\rm H}$>0 at the envelope scale, the opposite to other size distributions \\
$\ddagger$~H$^-$O: Hall+Ohmic model with anti-aligned configuration ($\bmath{\Omega \cdot B}<0$) \\
$\ddagger$~H$^+$O: Hall+Ohmic model with aligned configuration ($\bmath{\Omega \cdot B}>0$) \\
$\ddagger$~AO: AD+Ohmic model \\
$\ddagger$~Slw: model with slow initial core rotation \\
$\ddagger$~HO-NoRot: Hall+Ohmic model with zero initial core rotation \\
$\ast$~FHSC: first hydrostatic core \\
$\ast$~The $\uparrow$ or $\downarrow$ symbol indicates that the disc radius is growing or shrinking, repectively \\
$\ast$~The $^\circlearrowright$ symbol indicates that the disc is counter-rotating with respect to the initial core rotation \\
%\end{minipage}
\end{table}

\begin{table}
%\begin{minipage}{80mm}
\caption{Model Parameters for weak B-field $B_0\approx21.3~\mu$G ($\lambda$$\sim$4.8)}
\label{Tab:model2}
\resizebox{1.1\columnwidth}{!}{
\begin{tabular}{lcccc}
\hline\hline
Model & Grain Size & $\zeta_0^{\rm H_2}$ & $\beta_{\rm rot}$ & Radius \& Morphology \\
& Dist. & (10$^{-17}$~s$^{-1}$) & & (AU) \\
\hline
4.8MRN-H$^-$O & MRN & 1 & 0.1 & $\sim${\bf 7}$\downarrow$<2 \\
4.8MRN-H$^+$O & MRN & 1 & 0.1 & $\sim${\bf 12}$\downarrow$<2 \\
\hline
4.8min1-H$^-$O & min1 & 1 & 0.1 & $\uparrow$$\sim${\bf 50}$\downarrow$$\lesssim${\bf 15} \\
4.8min1-H$^+$O & min1 & 1 & 0.1 & $\sim${\bf 20}$\downarrow$<2 $\Rightarrow$ $\lesssim${\bf 20}$^\circlearrowright$$\downarrow$<{\bf 10}$^\circlearrowright$ \\
\hline
4.8opt3-H$^-$O & opt3 & 1 & 0.1 & $\uparrow${\bf 40--50}$\downarrow$$\lesssim${\bf 16} \\
4.8opt3-H$^+$O & opt3 & 1 & 0.1 & $\sim${\bf 25}$\downarrow$<2 $\Rightarrow$ {\bf 20--30}$^\circlearrowright$$\downarrow$<{\bf 10}$^\circlearrowright$ \\
4.8opt3-H$^-$O-Slw & opt3 & 1 & 0.025 & $\uparrow${\bf 40--50}$\downarrow$$\lesssim${\bf 18} \\
4.8opt3-H$^+$O-Slw & opt3 & 1 & 0.025 & $\downarrow$<2 $\Rightarrow$ {\bf 20--30}$^\circlearrowright$$\downarrow$<{\bf 10}$^\circlearrowright$ \\
4.8opt3-HO-NoRot & opt3 & 1 & 0 & $\uparrow$$\sim${\bf 40}$^\circlearrowright$$\downarrow${\bf 16}$^\circlearrowright$ \\
\hline
4.8trMRN-AO & trMRN & 1 & 0.1 & $\gtrsim${\bf 50}$\uparrow$ (Disc+Spiral/Ring) \\
4.8trMRN-H$^-$O & trMRN & 1 & 0.1 & $\uparrow${\bf 30--40}$\downarrow$<{\bf 10} \\
4.8trMRN-H$^+$O & trMRN & 1 & 0.1 & $\gtrsim${\bf 20}$\downarrow$<2 $\Rightarrow$ {\bf 20--30}$^\circlearrowright$$\downarrow$<{\bf 10}$^\circlearrowright$ \\
\hline
4.8LG-H$^-$O & LG & 1 & 0.1 & <2 \\
4.8LG-H$^+$O & LG & 1 & 0.1 & $\sim${\bf 10}$\downarrow$<2 \\
\hline
4.8CR10opt3-H$^-$O & opt3 & 10 & 0.1 & $\uparrow$$\sim${\bf 10}$\downarrow$$\lesssim$8 \\
4.8CR10opt3-H$^+$O & opt3 & 10 & 0.1 & $\uparrow$$\sim${\bf 15}$\downarrow$<2 $\Rightarrow$ $\lesssim$7$^\circlearrowright$ \\
\hline\hline
\end{tabular}
}
%\end{minipage}
\end{table}

\begin{table}
%\begin{minipage}{80mm}
\caption{Model Parameters for very weak B-field $B_0\approx10.6~\mu$G ($\lambda$$\sim$9.6)}
\label{Tab:model3}
\resizebox{1.1\columnwidth}{!}{
\begin{tabular}{lcccc}
\hline\hline
Model & Grain Size & $\zeta_0^{\rm H_2}$ & $\beta_{\rm rot}$ & Radius \& Morphology \\
& Dist. & (10$^{-17}$~s$^{-1}$) & & (AU) \\
\hline
9.6opt3-H$^-$O & opt3 & 1 & 0.1 & $\uparrow${\bf 40--50}$\downarrow$$\lesssim${\bf 16} \\
9.6opt3-H$^+$O & opt3 & 1 & 0.1 & >{\bf 50}$\uparrow$ (Disc+Spiral/Ring) \\
\hline\hline
\end{tabular}
}
\end{table}

\section{Simulation Results}
\label{Chap.SimulResult}

As summarized in Table~\ref{Tab:model1}--\ref{Tab:model2}, protostellar disc 
formation by Hall effect alone is sensitive to microphysics, especially the grain 
size distribution and cosmic-ionization rate. Similar to the case of AD 
\citep{Zhao+2016,Zhao+2018a}, disc formation is also suppressed in models with 
the standard MRN size distribution in which a large population of VSGs dominates 
the fluid conductivity and suppresses the Hall diffusivity. However, unlike the 
LG (single-sized 1~$\mu$m grain) models in the AD result \citep{Zhao+2016}, 
suppression of disc formation in LG models here is more severe than the MRN 
models, which is consistent with the values of Hall diffusivity found in 
\citet[][see their Fig.~5]{Zhao+2018b}.

In comparison to the existing work \citep{Tsukamoto+2015b,Wurster+2016} 
claiming that disc formation is enabled in the anti-aligned configuration 
but suppressed in the aligned configuration, we instead find no such bimodality 
when following the disc-envelope system into the main accretion phase. 
In the anti-aligned configuration ($\bmath{\Omega \cdot B}$<0), the initial 
discs ($\sim$30--50~AU) formed by Hall effect have only its inner region 
($\lesssim$10--20~AU) being long-lived RSDs while the outer region being 
rotationally supported Hall current sheets. %(termed as RSHCS below). 
In the aligned configuration ($\bmath{\Omega \cdot B}$>0), the initial disc 
suppression is followed by the formation of $\sim$20--40~AU counter-rotating 
discs that later evolve similarly to that in the anti-aligned case, with only 
the inner $\lesssim$10~AU RSD being long-lived.

\subsection{Inefficient Hall Effect in the MRN \& LG Models}
\label{S.MRN_LG}

\begin{figure*}
\begin{tabular}{ccc}
& \includegraphics[width=1.4\columnwidth]{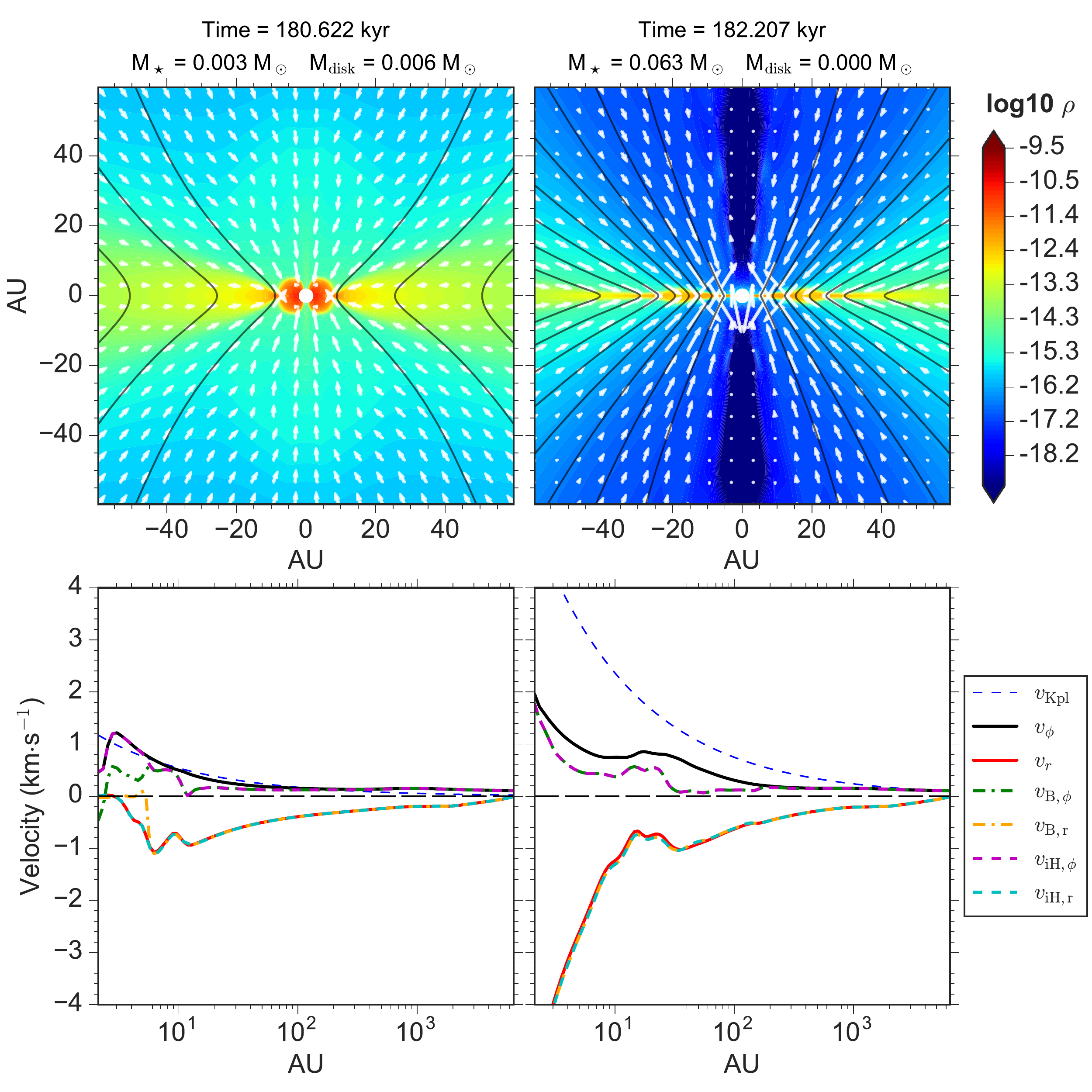} & 
\includegraphics[width=0.7\columnwidth]{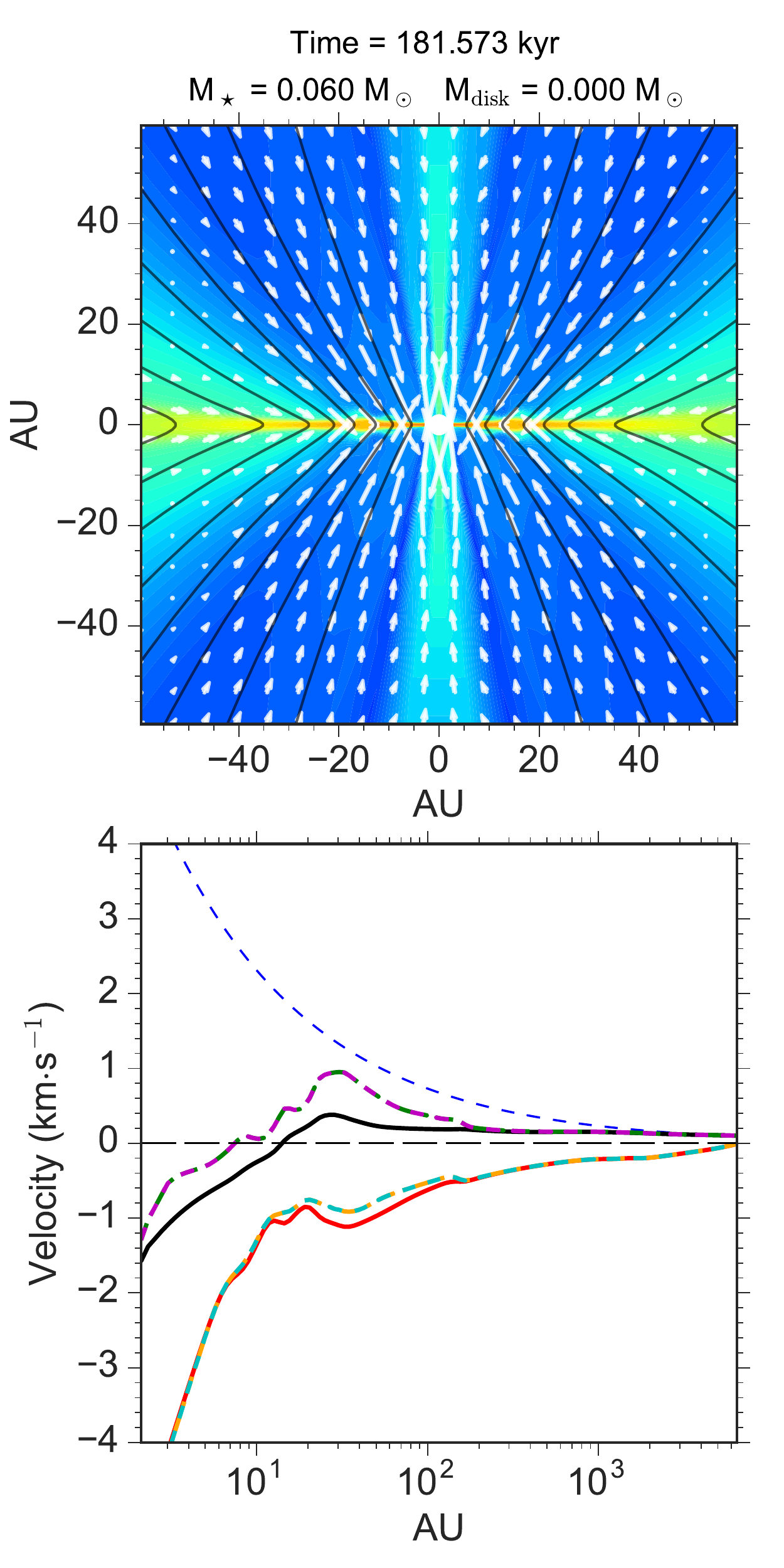}\\
& (a) Anti-aligned: $\bmath{\Omega \cdot B}<0$ & (b) Aligned: $\bmath{\Omega \cdot B}>0$\\
\end{tabular}
\caption{Mass density distribution (top) and velocity profile along the equator 
(bottom) for model 2.4MRN-H$^-$O (left \& middle panels) and model 2.4MRN-H$^+$O 
(right panel). White arrows and black lines in the top panel are the velocity field 
vectors and magnetic field lines, respectively. The disc mass is estimated 
according to the criteria in Appendix.~\ref{App.A}.}
\label{Fig:2.4MRN}
\end{figure*}
\begin{figure}
\includegraphics[width=\columnwidth]{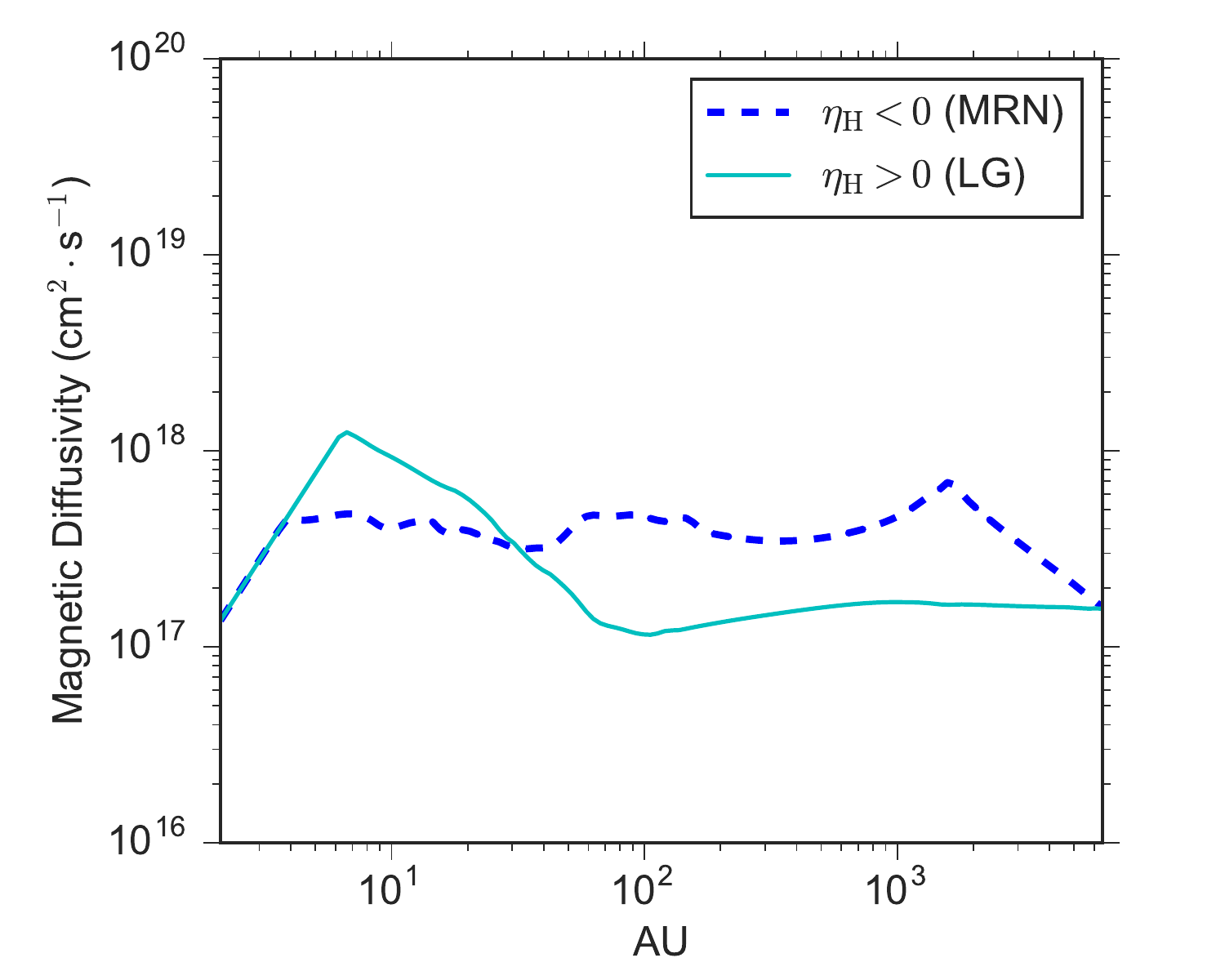}
\caption{Radial profile of Hall diffusivity along the equator for the 2.4MRN-H$^-$O 
at $t=182.207$~kyr and the 2.4LG-H$^+$O model at $t=181.890$~kyr, respectively, 
when no disc is present and the stellar mass reaches $\sim$0.06~M\sun. Note that 
$\eta_{\rm H}$ is mostly positive in the LG models, hence the aligned LG models 
behave similarly to the anti-aligned MRN models.}
\label{Fig:2.4comp_etas}
\end{figure}

We first show that Hall effect is inefficient with either the standard MRN 
size distribution or singly-sized large grains (LG), which are adopted by 
\citet{Li+2011}. As shown in Fig.~\ref{Fig:2.4MRN}, 
even with an anti-aligned configuration ($\bmath{\Omega \cdot B}<0$) which has 
previously shown to promote disc formation, the 2.4MRN-H$^-$O model show no 
rotationally supported disc larger than 2~AU (inner boundary) after the first 
core stage. The azimuthal Hall drift does cause the magnetic field to move 
somewhat slower azimuthally than the gas rotation, slightly reducing the 
azimuthal bending of magnetic field lines. However, such an azimuthal drift 
is relatively small due to the low Hall diffusivity $\eta_{\rm H}$ 
at 10--100~AU scale ($\sim$a few 10$^{17}$~cm$^2$~s$^{-1}$; see 
Fig.~\ref{Fig:2.4comp_etas}). Thus, gas falls in with large $\varv_r$ 
along the pseudo-disc in the absence of sufficient rotation support, 
dragging the field lines inward. The radial velocity of either 
the charged species that dominates the Hall diffusivity ($\varv_{{\rm iH},r}$) 
or the effective radial velocity of the magnetic field ($\varv_{{\rm B},r}$) are 
almost indistinguishable from the gas infall velocity $\varv_r$, which is also a 
sign of insignificant azimuthal bending (towards either +$\phi$ or -$\phi$) of 
magnetic fields according to Eq.~\ref{Eq:v_Hr}. 

In the aligned case with MRN size distribution (model 2.4MRN-H$^+$O), 
disc formation is also suppressed with an even smaller gas rotation $\varv_r$ 
within the inner $\sim$100~AU compared to the anti-aligned case at a similar stage 
(in terms of total mass in the central region). As expected from Eq.~\ref{Eq:v_Hphi}, 
the azimuthal Hall drift goes along the direction of the original gas rotation 
(+$\phi$), causing the magnetic field to surpass the azimuthal gas motion along 
+$\phi$. At $\sim$10--100~AU scale, the enhanced field bending towards +$\phi$ 
strengthens the magnetic tension force towards -$\phi$ and slows down the gas rotation. 
On the other hand, the azimuthal field bending induces a radially-outward Hall drift 
of the magnetic field, which tends to curb the azimuthal Hall effect 
(\S~\ref{S.HallDrift}). Inside $\lesssim$10~AU, the magnetic torque not only 
brakes the gas rotation, but spins the gas in the opposite direction (-$\phi$) 
as well, creating a counter-rotating region of a few AU but with insufficient 
rotational velocity to form a RSD.

Similar to the MRN models, the LG models in Table~\ref{Tab:model1}--\ref{Tab:model2} 
also fail to form RSDs, producing only transient first-core-like structures that last 
less than 1--2~kyr. Besides, such transient structures appear to be smaller 
in size than the MRN models, owing to a lower Hall diffusivity 
($\eta_{\rm H}$$\sim$10$^{17}$~cm$^2$~s$^{-1}$ as shown in Fig.~\ref{Fig:2.4comp_etas}) 
throughout the core, especially between tens AU to 100~AU scale where Hall effect 
is the most relevant. Note that, unlike other cases with sub-micron grain, 
$\eta_{\rm H}$ in the LG case ($a=1.0~\mu$m) is positive in the density range of 
the dense core \citep[$\gtrsim$10$^4$~cm$^{-3}$;][]{Zhao+2018b}. Therefore, the 
aligned configuration in the LG models is equivalent to the anti-aligned 
configuration in other models with sub-micron grains.

The result presented in this section is in agreement with \citet{Li+2011}, 
who find Hall effect to be inefficient in affecting disc formation, 
primarily due to their choice of two specific grain size distributions with 
relatively low Hall diffusivities. In comparison, recent work showing disc formation 
enabled by Hall effect \citep{Tsukamoto+2015b,Wurster+2016,Tsukamoto+2017,Wurster+2018} 
mostly adopt other size distributions (e.g., 0.1~$\mu$m) with enhanced Hall diffusivities.
As shown in Table~\ref{Tab:model1}--\ref{Tab:model2}, a slightly truncated MRN 
size distribution ($a_{\rm min} \gtrsim 0.01~\mu$m) indeed facilitates the 
initial disc formation enabled by Hall effect.

\subsection{Disc Formation \& Evolution with Enhanced Hall Diffusivity: 
Anti-Aligned Configuration}
\label{S.HallDisc-}

We now present the models adopting $a_{\rm min}=0.03~\mu$m in which Hall 
diffusivity $\eta_{\rm H}$ is enhanced by $\sim$1--2 orders of magnitude 
compared to that of MRN and LG grains, in the number density range of 
10$^9$--10$^{12}$~cm$^{-3}$ \citep{Zhao+2018b}. With such an enhanced $\eta_{\rm H}$, 
Hall effect strongly dominates the gas dynamics and the evolution of magnetic 
fields in the inner envelope. We first demonstrate the case of 
anti-aligned configuration ($\bmath{\Omega \cdot B}$<0).

\subsubsection{Disc Formation and Growth at Early Times}
\label{S.HallGrowth}

\begin{figure*}
\centerline{\includegraphics[width=1.17\textwidth]{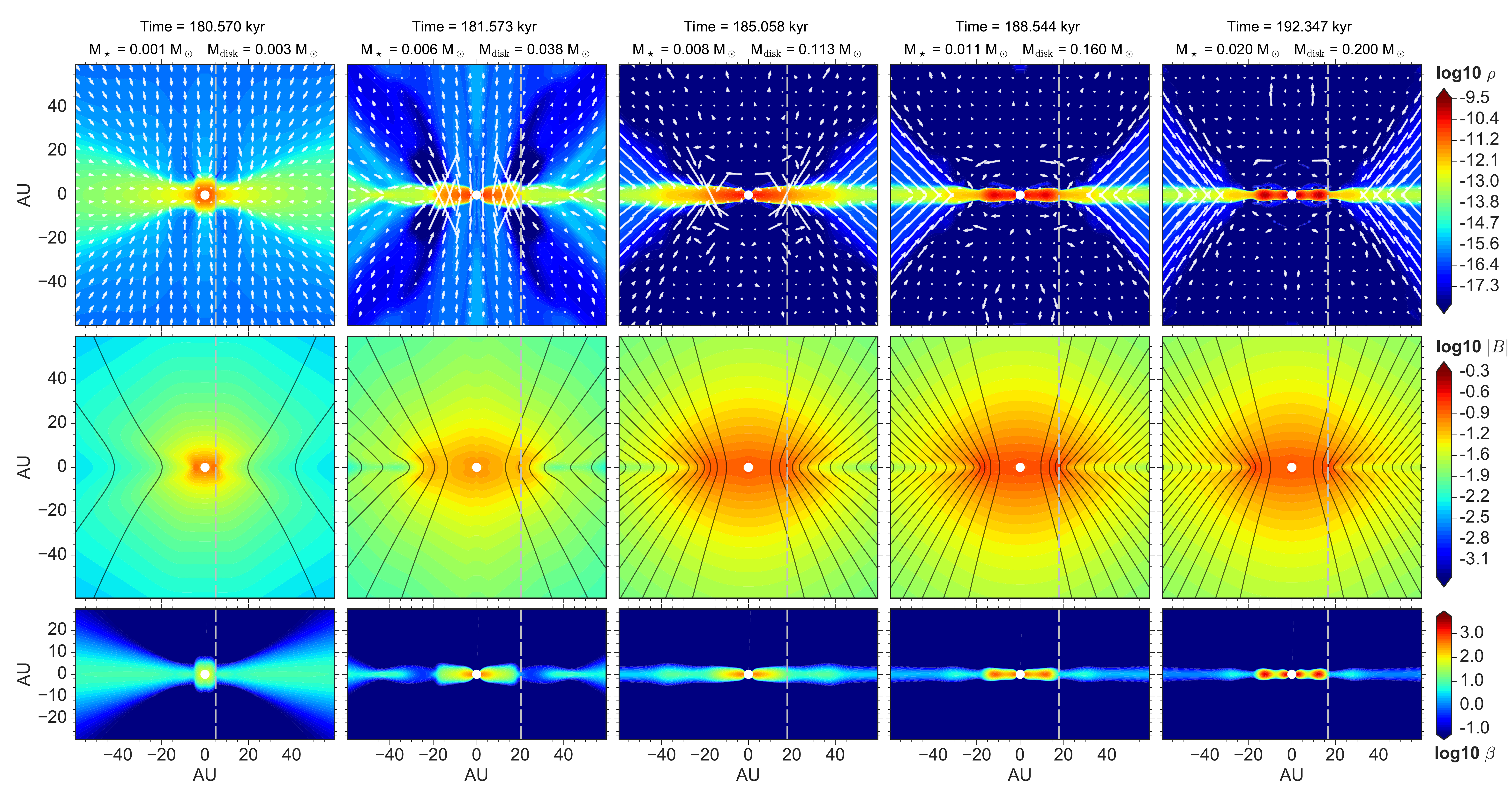}}
\centerline{\includegraphics[width=1.17\textwidth]{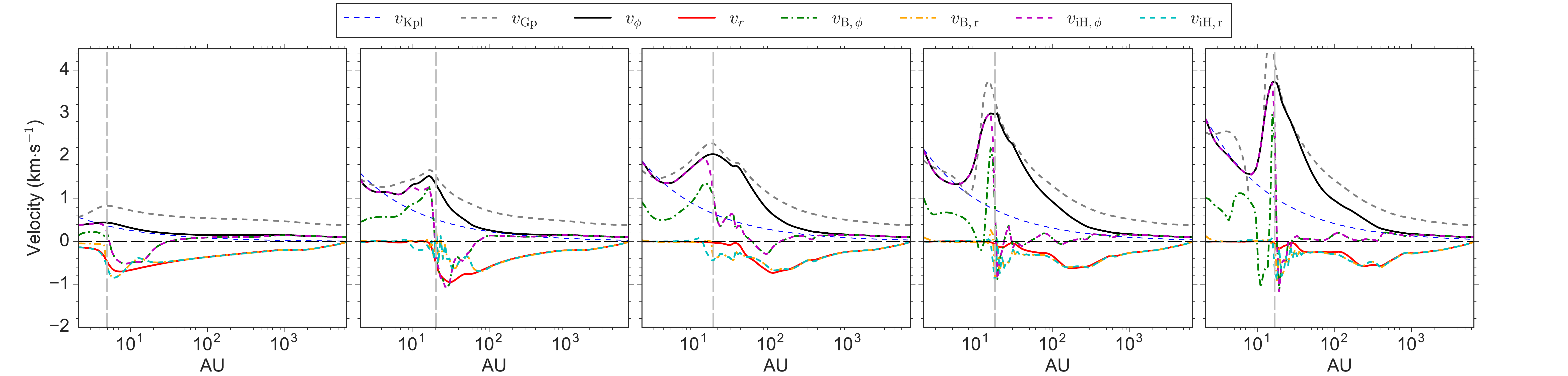}}
\caption{Evolution of disc in the anti-aligned model 2.4opt3-H$^-$O. First row: 
logarithmic distribution of mass density along with velocity field vectors (white arrows). 
Second row: logarithmic distribution of total magnetic field strength $|B|$ along 
with magnetic field lines (black solid lines). Third row: logarithmic distribution 
of plasma-$\beta$. Fourth row: velocity profile along the equator. The vertical 
silver line (dashed) approximately marks the edge of the inner RSD. 
Note that the relatively large mass ratio of disc to star is mainly because 
of the 2D axisymmetric set-up that limits the channel of mass accretion to the 
central stellar object \citep[see also][]{Zhao+2016}; breaking of axisymmetry in 
3D would significantly lower the mass ratio of disc to star \citep{Zhao+2018a}.}
\label{Fig:2.4opt3-}
\end{figure*}
\begin{figure*}
\includegraphics[width=\textwidth]{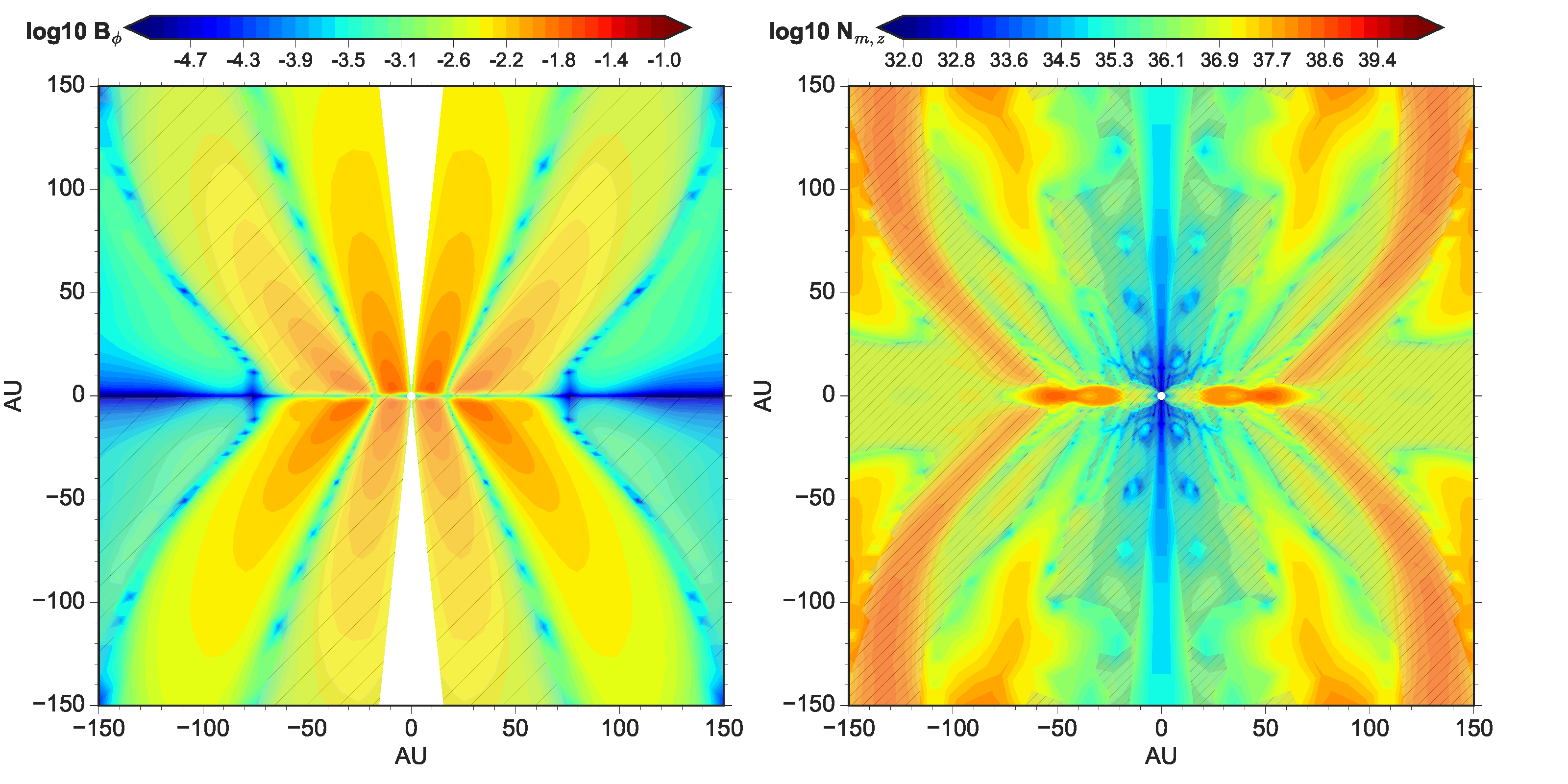}
\caption{Logarithmic distribution of azimuthal magnetic field $B_\phi$ (left panel) and 
magnetic torque $N_{{\rm m},z}$ (right panel) at $t=181.573$~kyr of the anti-aligned 
model 2.4opt3-H$^-$O. $B_\phi$ is positive (along +$\phi$) in the unshaded region and 
negative (along -$\phi$) in the shaded region. Similarly, regions of negative magnetic 
torque (along -$z$) are shown as shaded.}
\label{Fig:BphTq_early}
\end{figure*}

Fig~\ref{Fig:2.4opt3-} shows the time evolution of the inner 60~AU region of 
model 2.4opt3-H$^-$O, in which an initial disc that formed after the first 
core stage grows within the first $\sim$4~kyr to 40--50~AU radius 
\citep[similar to][]{Tsukamoto+2015b,Wurster+2016}. 
The formation of the initial disc is caused by over-bending the magnetic 
field lines towards -$\phi$ direction (the middle field line of the upper 
panel of Fig.~\ref{Fig:sketch-B}), which generates a pure spin-up torque that 
accelerates the gas rotation along +$\phi$-direction. 
For example, at $t=181.573$~kyr (Fig.~\ref{Fig:BphTq_early}) 
when the disc is in its early growth phase, the azimuthal magnetic field 
$B_\phi$ flips sign (bends towards -$\phi$) between $\sim$20--70~AU, which 
naturally produces positive torques in the same region along the equator 
(right panel of Fig.~\ref{Fig:BphTq_early}). The region of field over-bending 
almost coincides with the region of negative $\varv_{\rm B,\phi}$ in 
Fig.~\ref{Fig:2.4opt3-}, also indicating that the original azimuthal field 
bending along +$\phi$ is not as severe and can be easily bent towards -$\phi$. 
Accordingly, such a $B_\phi$ geometry also induces a large radial Hall current 
that drifts the magnetic field outward along the equator, with a drift velocity 
$\varv_{{\rm H},r} \sim +0.5$~km~s$^{-1}$ (the difference between 
$\varv_r$ and $\varv_{{\rm iH},r}$). Therefore, both the spin-up torque 
operating azimuthally and the outward drift of magnetic fields radially 
aid the disc formation and growth in the early phase. Note that in the 
inner $\sim$20~AU disc, where gas rotation becomes fast, the azimuthal magnetic 
field $B_\phi$ is flipped back again to +$\phi$-direction and magnetic torque 
becomes weakly negative (at these early times).

\begin{figure}
\centerline{\includegraphics[width=1.1\columnwidth]{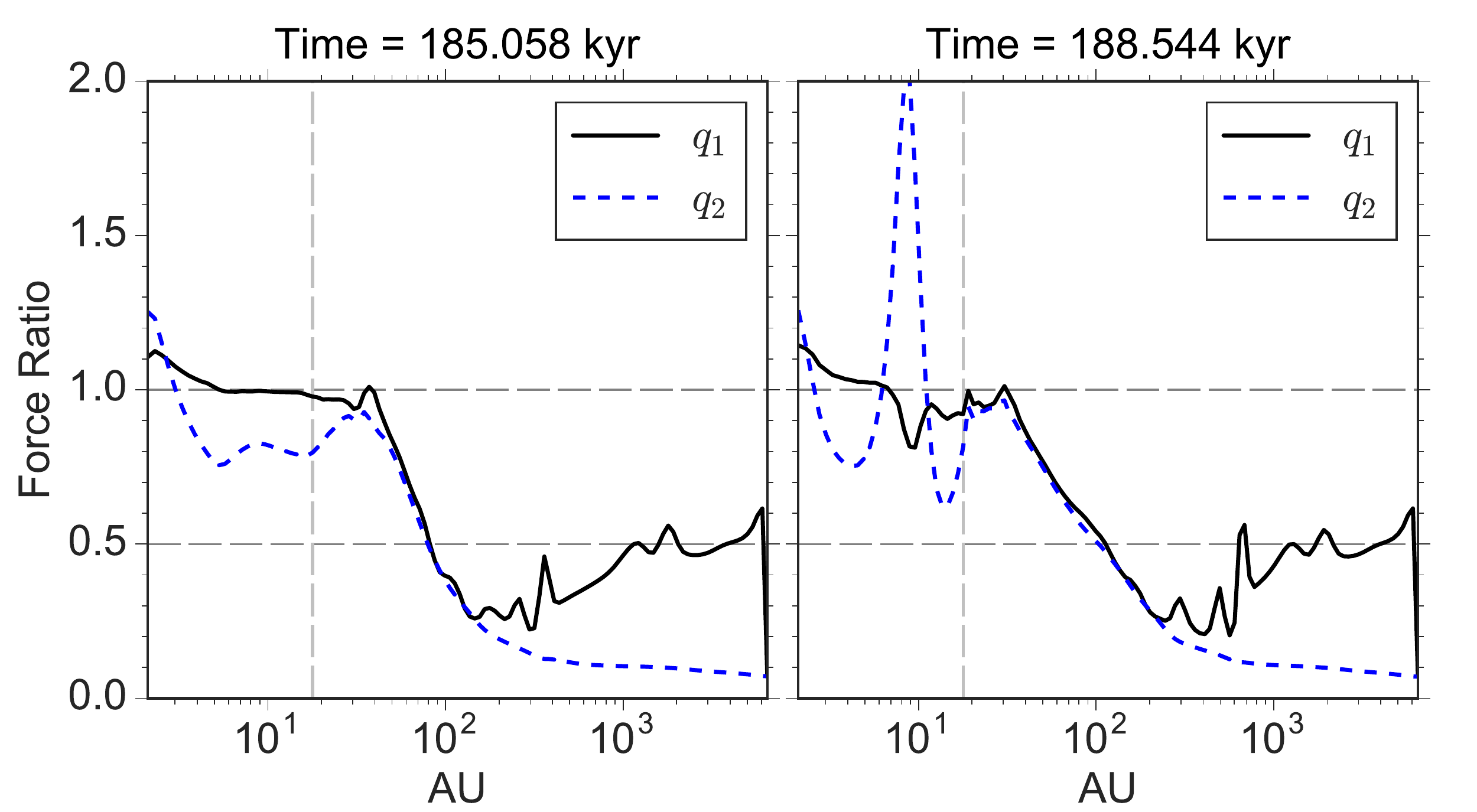}}
\caption{Ratio of the dominant forces along the equator. Note that the inner 
RSD ($\lesssim$16~AU) at later times (right panel) becomes self-gravitating, 
hence the ratios $q_1$ and $q_2$ may no longer be accurate.}
\label{Fig:2.4comp_forces}
\end{figure}

Despite the super-Keplerian rotation speed ($\varv_\phi>\varv_{\rm Kpl}$) due 
to a low protostellar mass, the disc is gravitationally bound with 
$\varv_\phi\lesssim\varv_{\rm Gp}$, where 
\begin{equation}
\varv_{\rm Gp} = \sqrt{r {\partial \Phi \over \partial r}}~, \hspace{35pt}\Phi(\infty)\rightarrow0
\end{equation}
and $\Phi<0$ is the gravitational potential (stellar potential and gas self-gravity) 
at distance $r$. Fig~\ref{Fig:2.4opt3-} shows that the discs in the early phase 
after forming the first core (e.g., $t=181.573$~kyr and $t=185.058$~kyr) are 
rotating with a speed close to $\varv_{\rm Gp}$. To confirm the discs at early 
times are mostly rotationally supported, we examine the ratio of the dominant forces 
along the equator. Similar to \citet{Tsukamoto+2015b} and \citet{Wurster+2018}, 
we define the ratio of the sum of the centrifugal and pressure gradient forces 
to the gravitational force as,
\begin{equation}
q_1 = \left|{{{\varv_\phi^2 \over r} + {1 \over \rho}{\partial P \over \partial r}} \over {\partial \Phi \over \partial r}}\right|~,
\end{equation}
and the ratio of the centrifugal force to the gravitational force as,
\begin{equation}
q_2 = \left|{{\varv_\phi^2 \over r} \over {\partial \Phi \over \partial r}}\right|~.
\end{equation}
The left panel of Fig.~\ref{Fig:2.4comp_forces} shows that, at $t=185.058$~kyr 
when the disc radius reaches its maximum during the evolution, the inner 
$\lesssim$40--50~AU region is supported by rotation and thermal pressure against 
the gravitational collapse. Indeed, the sum of centrifugal and pressure gradient 
forces approximately balances the gravitational force ($q_1$$\sim$1), and the 
main contributor against gravity is the centrifugal support ($q_2$$\gtrsim$0.8). 

\subsubsection{Disc Evolution \& RSHCS at Later Times}
\label{S.HallSheet}

After $t=185.058$~kyr, the bulk disc gradually divides itself into two distinct 
partitions: (1) a long-lived inner RSD of <20~AU radius and (2) an outer Hall 
current sheet (between $\sim$20--40~AU) that flattens vertically over a period of 
5--10~kyr. As marked in Fig.~\ref{Fig:2.4opt3-}, we separate the two partitions 
by a vertical line, near which the magnetic field strength peaks and 
$\eta_{\rm H}$ changes sign (for models with sub-micron grains). Interior to 
the vertical line, $\eta_{\rm O}$ increases rapidly and the magnetic field 
strength saturates. Therefore, the vertical line approximately defines the edge 
of the inner RSD.

Although at later times the outer $\sim$20--40~AU partition remains rotationally 
supported along the radial direction (right panel of Fig.~\ref{Fig:2.4comp_forces})
due to efficient azimuthal Hall drift (fourth row of Fig.~\ref{Fig:2.4opt3-}), 
it flattens vertically over time as the ``fan-like'' magnetic field lines 
remains highly pinched. In fact, despite the large rotational velocity 
($\varv_\phi\rightarrow\varv_{\rm Gp}$), the outer partition is essentially 
a variant of the pseudo-disc structure 
\citep{GalliShu1993a,GalliShu1993b,Allen+2003a,Allen+2003b}, 
with envelope gas sliding freely ($\sim$2~km~s$^{-1}$) along the pinched 
field lines and flattening the outer partition. However, the highly pinched field 
configuration threading the outer partition is caused by the radially inward Hall 
drift of magnetic fields, instead of the dragging of magnetic field lines by 
gas infall along an usual pseudo-disc. As shown in Fig.~\ref{Fig:2.4opt3-}, 
$\varv_{{\rm H},r}$ between $\sim$20--40~AU is about 0.5--1~km~s$^{-1}$ 
while gas infall speed $\varv_r$ there is nearly vanishing. Furthermore, 
the plasma-$\beta$ ($\equiv {P_{\rm th} \over P_{\rm B}}$, where 
$P_{\rm th}$ is the thermal pressure and $P_{\rm B}$ the magnetic pressure) 
in the outer partition also decreases by 1 order of magnitude from a few 10$^1$ 
at early times to around unity (third row in Fig.~\ref{Fig:2.4opt3-}). 
To avoid confusion with usual pseudo-discs. we term such a pseudo-disc with 
rotational support as ``{\it rotationally supported Hall current sheet}'' 
({\bf RSHCS}). 

\begin{figure*}
\includegraphics[width=\textwidth]{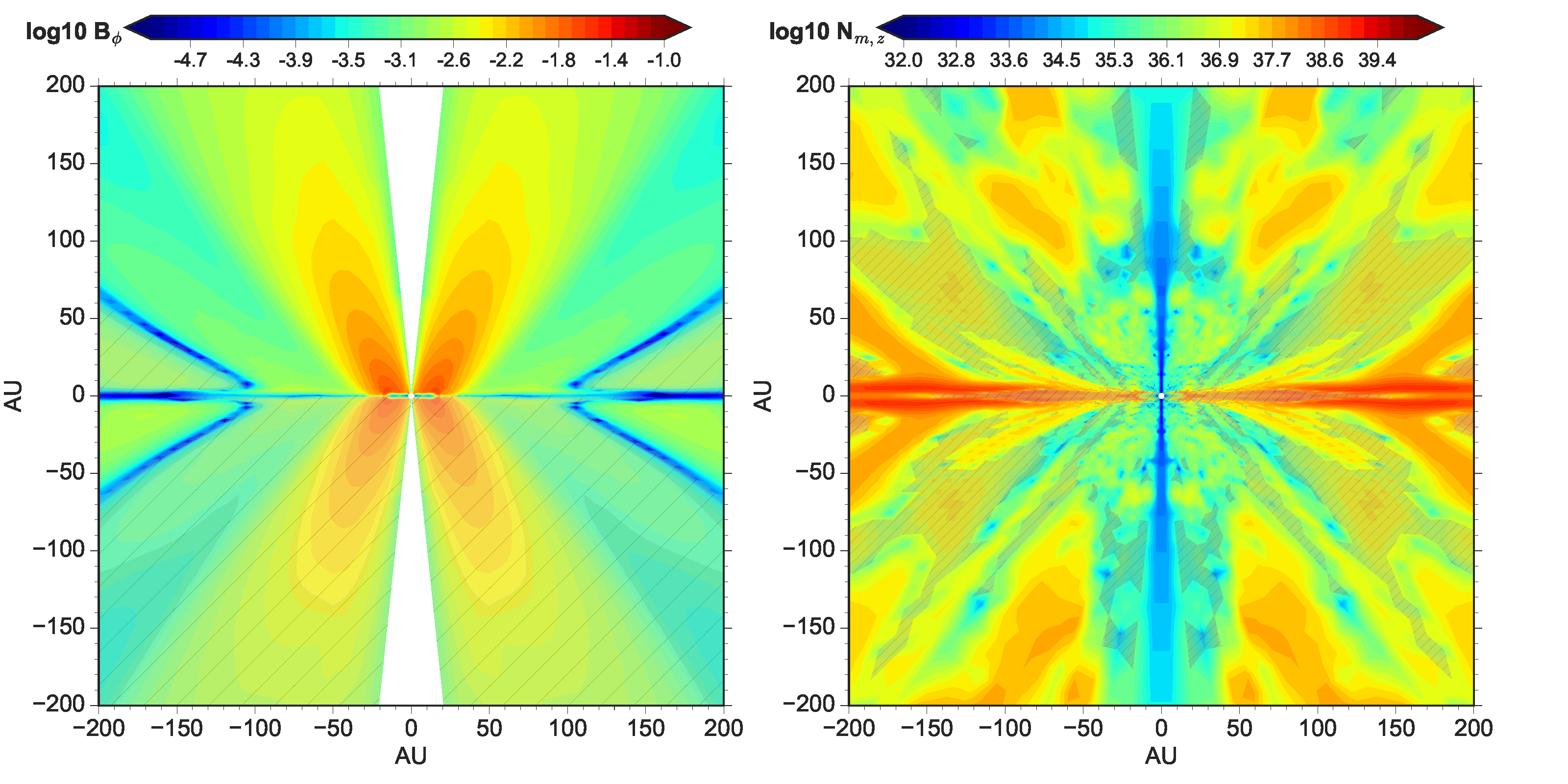}
\caption{Same as Fig~\ref{Fig:BphTq_early}, but for a later time $t=188.544$~kyr 
of model 2.4opt3-H$^-$O. The shaded regions are negative $B_\phi$ (left) and 
negative magnetic torque (right), respectively.}
\label{Fig:BphTq_late}
\end{figure*}

\begin{figure*}
\includegraphics[width=\textwidth]{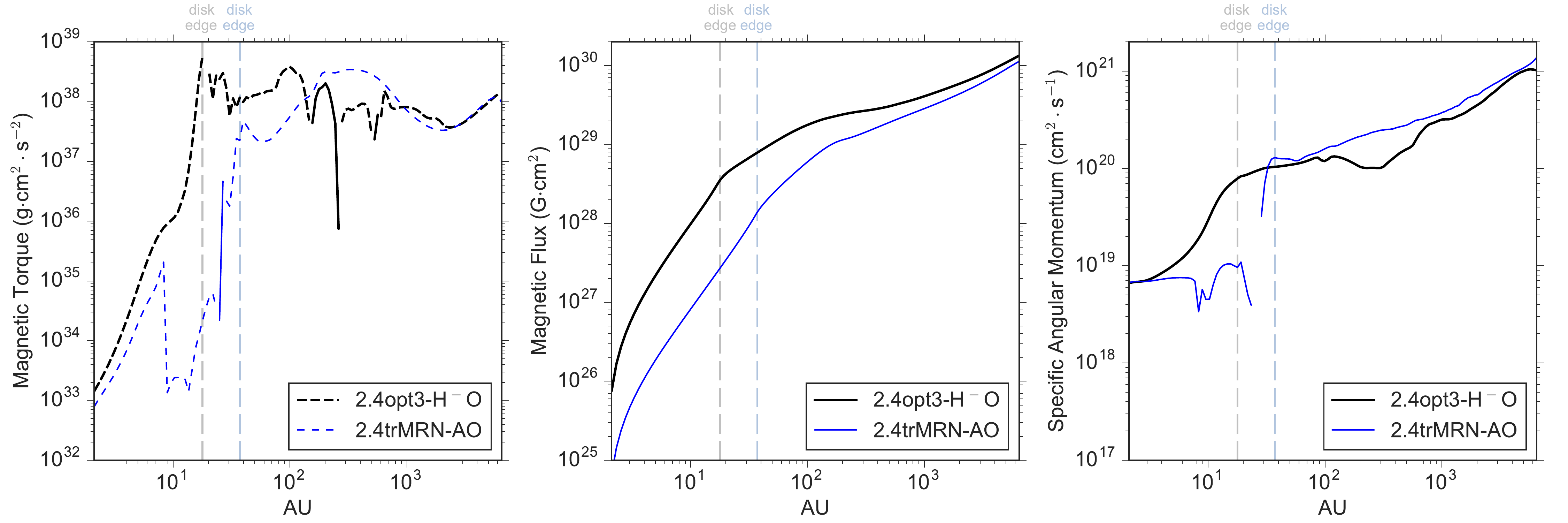}
\caption{Profile of magnetic torque along the equator (left panel), magnetic flux 
threading through the equatorial plane (middle panel), and specific angular momentum 
in concentric cylindrical shells (right panel) for model 2.4opt3-H$^-$O at $t=188.544$~kyr 
and model 2.4trMRN-AO at $t=167.947$~kyr, respectively, when the total mass of star and 
disc reaches a similar value 0.17~M\sun. The comparison is mainly for the regions 
outside the disc edge (vertical dashed lines). The solid curves in the left panel 
denote positive magnetic torques (along +$z$).}
\label{Fig:2.4comp_torq}
\end{figure*}

In the RSHCS (i.e., the outer partition), though large azimuthal Hall drift 
efficiently weakens the azimuthal bending of magnetic field across the equator, 
yet the magnetic braking torque remains strong because of the accumulation of 
poloidal field $B_z$ (Eq.~\ref{Eq:Lz}) in the RSHCS by the radially inward 
Hall drift. At $t=188.544$~kyr, magnetic braking torque is strongly negative 
(towards -$z$ in Fig.~\ref{Fig:BphTq_late}), with 
$N_{{\rm m},z}$$\sim$10$^{38}$~g~cm$^2$~s$^{-1}$ in the inner 
$\lesssim$120~AU equatorial region (see also Fig.~\ref{Fig:2.4comp_torq}).
Basically, the gas rotation that drags $B_\phi$ towards +$\phi$ direction 
becomes fast in the inner region (Fig.~\ref{Fig:2.4opt3-}), hindering the 
over-bending of $B_\phi$ towards -$\phi$ across the equator by the azimuthal Hall 
drift (left panel of Fig.~\ref{Fig:BphTq_late}; see also the sketch in 
Fig.~\ref{Fig:sketch-B}), even when the effective azimuthal velocity of the magnetic 
field $\varv_{{\rm B},\phi}$ (or similarly $\varv_{{\rm iH},\phi}$) 
in some locations becomes negative, i.e., points to -$\phi$. 
Positive magnetic torque only appears in the outer region ($\sim$120--250~AU) 
where gas rotation along +$\phi$ is slow and azimuthal Hall drift towards -$\phi$ 
is sufficient to over-bend $B_\phi$. 

In other words, were it not for the efficient azimuthal Hall drift, the 
magnetic braking torque in the RSHCS would have been even stronger with the 
same magnitude of poloidal magnetic field $B_z$; in contrast, efficient 
radially outward drift of magnetic fields such as AD would easily resolve 
such a tension. In Fig.~\ref{Fig:2.4comp_torq}, we compare the magnetic 
torques of the 2.4opt3-H$^-$O model to the 2.4trMRN-AO model, at a similar 
evolutionary stage. Recall from Table~\ref{Tab:model1} that the 2.4opt3-H$^-$O 
model is a Hall+Ohmic model with the most enhanced $\eta_{\rm H}$, whereas 
the 2.4trMRN-AO is an AD+Ohmic model with the most enhanced $\eta_{\rm AD}$ 
\citep{Zhao+2018b}.\footnote{The AD+Ohmic model forms an axisymmetric ring of 
30--40~AU \citep[Appendix.~\ref{App.A}; see also][]{Zhao+2016} that would 
actually be a grand design spiral structure in 3D \citep{Zhao+2018a}.} 
The magnetic torque in the AD+Ohmic model is lower by 1 order of magnitude 
in the inner $\lesssim$100~AU region than that in the Hall+Ohmic model, 
which is primarily a result of the difference in the (poloidal) magnetic 
field strength (Fig.~\ref{Fig:2.4opt3-} and Fig.~\ref{Fig:2.4tr-AO}). 
This also implies that the efficient azimuthal Hall drift is unable to 
fully offset the increase of magnetic torque by a stronger $B_z$. 
In contrast, magnetic decoupling in the AD+Ohmic model is already efficient 
at a broad scale (a few 100~AU to a few 1000~AU) in the envelope \citep{Zhao+2018a}, 
which notably reduces the amount of magnetic flux dragged into the inner 
envelope as well as the resulting value of $B_z$. In Fig.~\ref{Fig:2.4comp_torq}, 
the amount of magnetic flux throughout the envelope in the Hall+Ohmic model is 
a few times higher than in the AD+Ohmic model; the Hall+Ohmic model not only lacks 
an efficient mechanism to drift magnetic fields radially outward, but instead 
has radially inward drift of magnetic fields in the RSHCS to further increase 
the poloidal magnetic field and magnetic torque (see also \S~\ref{Dis.HallOsc} 
for further discussions).

\subsubsection{Thin Counter-Rotating Shells in Anti-Aligned Models \& Angular Momentum Conservation}
\label{S.CounterShell}

\begin{figure*}
\includegraphics[width=\textwidth]{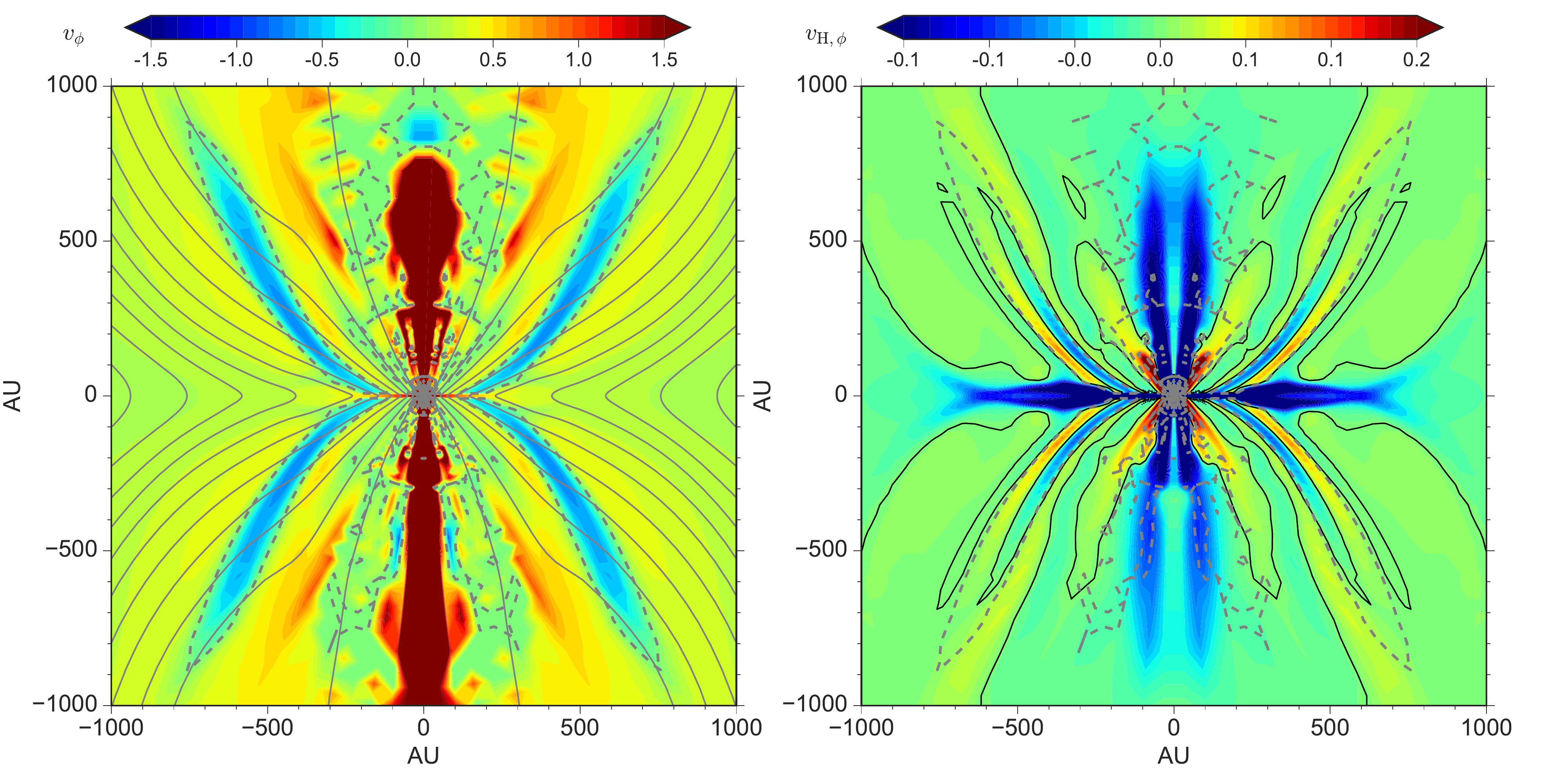}
\caption{Distribution of azimuthal velocity $\varv_\phi$ (left panel) and 
azimuthal Hall drift velocity $\varv_{{\rm H},\phi}$ (right panel) at 1000~AU 
scale envelope for model 2.4opt3-H$^-$O at $t=188.544$~kyr. Negative $\varv_\phi$ 
and $\varv_{{\rm H},\phi}$ values represent rotation and drift motions along -$\phi$ 
direction, respectively. 
Dashed contour lines (plotted in both left and right panel) mark positions with 
$\varv_\phi=0$, where the transition between positive and negative $\varv_\phi$ 
occurs. Similarly, the solid contour lines (only in the right panel) is for 
$\varv_{{\rm H},\phi}=0$. Grey solid lines trace the magnetic field lines.}
\label{Fig:VHph_late}
\end{figure*}

We confirm the existence of thin counter-rotating shells at 100--1000~AU scales 
in models of anti-aligned configuration, for example, the ``butterfly-shaped'' 
region in Fig.~\ref{Fig:VHph_late} (left panel) from model 2.4opt3-H$^-$O 
\citep[see also][]{Krasnopolsky+2011,Li+2011,Tsukamoto+2015b}. 
The shell with counter-rotation ($\varv_\phi<0$) coincides with the transition region 
where poloidal magnetic field lines change from a convex curve (bending towards the 
equator) to a concave curve for the upper quadrants (or from concave to convex 
for the lower quadrants). The convex magnetic curves around the counter-rotating shell 
induce a small azimuthal Hall drift of magnetic field along +$\phi$ (right panel of 
Fig.~\ref{Fig:VHph_late}), which strengthens the magnetic tension force along -$\phi$, 
slowing down the original gas rotation and even generating counter-rotation. 
However, the slight increase in azimuthal bending of magnetic field lines 
towards +$\phi$ induces a secondary Hall drift along the normal vector (pointing away 
from the equator) of the poloidal magnetic curve that in turn reduces the curvature 
of the convex magnetic curve. Along the ridge of the counter-rotating shell, where 
the local curvature of the poloidal magnetic curve as well as the azimuthal Hall 
drift both approaches 0, the counter-rotating velocity $\varv_\phi$ reaches a maximum. 
The azimuthal magnetic field also flips to bend towards -$\phi$, mostly following the 
azimuthal gas motion ($|\varv_{{\rm H},\phi}| < |\varv_\phi|$). Similarly, a secondary 
Hall drift pointing towards the equator is induced. Therefore, the Hall drift along 
the normal vector of the poloidal magnetic curve and along $\phi$ direction feedbacks 
and constrains each other. The magnetic field geometry alternates along both directions 
and leads to the counter-rotating shells. 

The counter-rotating shells contribute negative proportion to the total angular 
momentum budget, as shown in Fig.~\ref{Fig:2.4comp_torq} (right panel), where 
a dip along the specific angular momentum distribution clearly exists between 
100--1000~AU, in accordance with the scale of the counter-rotating shells. 
Inside 100~AU, the specific angular momentum of the Hall+Ohmic model increases 
again and becomes comparable to that of the AD+Ohmic model. Basically, the 
topology of magnetic field lines is regulated by Hall effect, extracting the 
angular momentum from the outer envelope and redistributing it in the inner 
envelope, which compensates for the strong magnetic braking near the RSHCS 
and sustains the inner RSD (<20~AU). 
Nonetheless, the total angular momentum is approximately conserved throughout the 
simulation, with a minor decrease by $\sim$5\% over a course of $\sim$10~kyr 
(e.g., from $1.08\times10^{54}$~g~cm$^2$~s$^{-1}$ at the first core stage 
to $1.03\times10^{54}$~g~cm$^2$~s$^{-1}$ at $\sim$190~kyr for the 
2.4opt3-H$^-$O model) due to the outflow boundary condition imposed on the inner 
and outer boundaries of the simulation box \citep[similar to][]{Krasnopolsky+2010}. 
Note that the violation of angular momentum conservation found in \citet{Marchand+2018} 
is likely originated from the numerical scheme they adopted for Hall effect, as well as 
from the Cartesian grid structure. Indeed, their follow-up study \citep{Marchand+2019} 
shows that removing the whistler wave speed in the Riemann solver for non-magnetic 
variables can reduce the spurious generation of angular momentum by one order of 
magnitude and improve the angular momentum conservation.
%The narrow region near $\sim$200~AU in the Hall+Ohmic model with positive 
%torques coincides with the region of ``over-bended'' $B_\phi$ as shown in 
%Fig.~\ref{Fig:BphTq_late} above.

\subsubsection{Saturation of Magnetic Field Strength Inside RSD by Ohmic Dissipation}
\label{S.Ohmic}

\begin{figure*}
\includegraphics[width=1.0\textwidth]{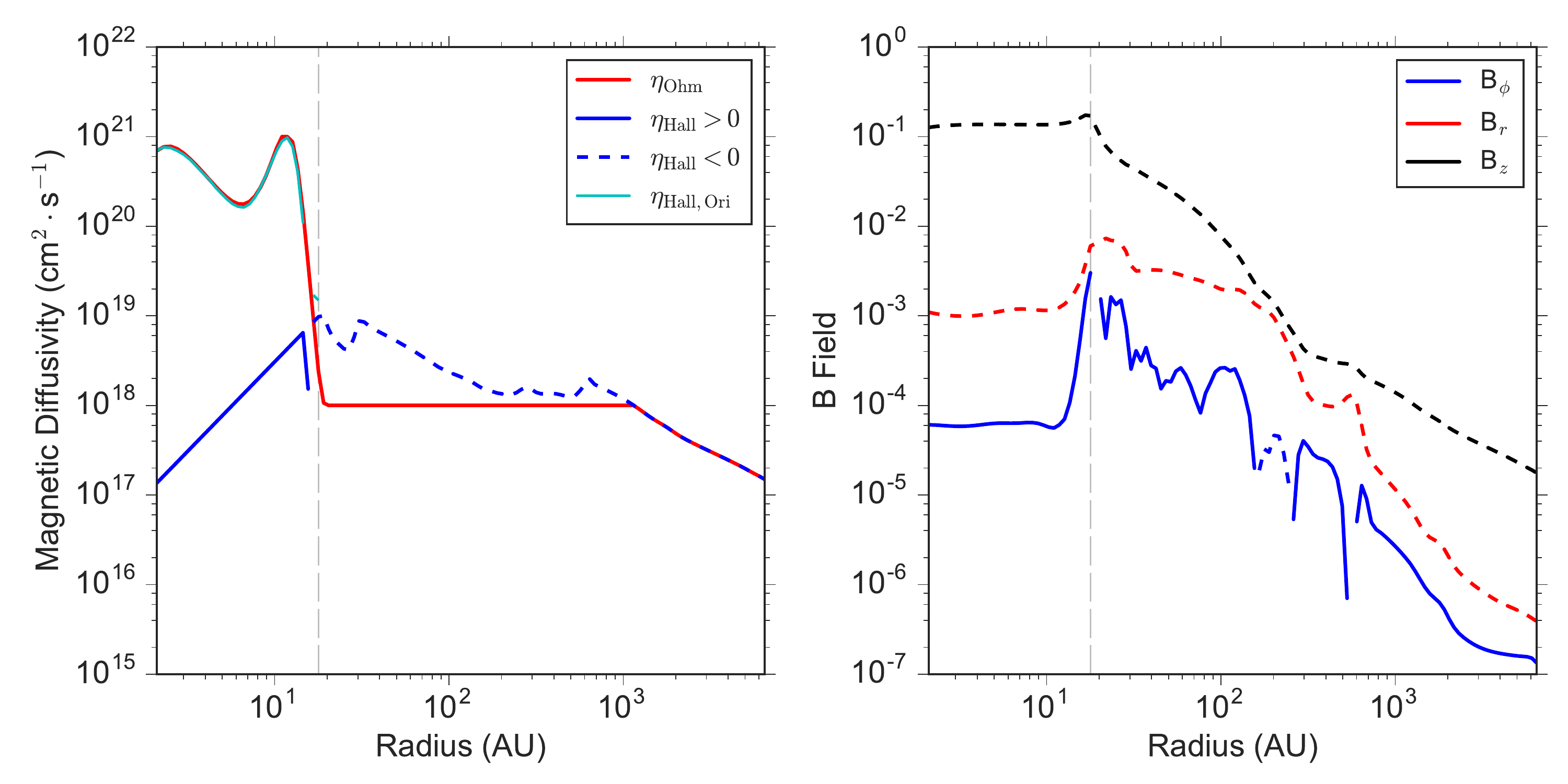}
\caption{Profile of magnetic diffusivities (left panel) and magnetic field strengths 
(right panel) along the equator for model 2.4opt3-H$^-$ at $t=188.544$~kyr. 
$\eta_{\rm H,Ori}$ shows the original Hall diffusivity unaffected by the Hall 
d$t$ floor.}
\label{Fig:EtaBc_late}
\end{figure*}

In the inner RSD, the three components of the magnetic field saturate near the mid-plane,
with poloidal $B_z$ ($\sim$0.1~G) being the dominant component and $B_r$ or $B_\phi$ 
2--3 orders of magnitude smaller than $B_z$. It is a direct outcome of the highly 
efficient Ohmic diffusion in the inner RSD where $\eta_{\rm O}$ rises to 
10$^{20}$--10$^{21}$~cm$^2$~s$^{-1}$ and the effective velocities of the 
magnetic field lines are suppressed (Fig.~\ref{Fig:2.4opt3-}). 
In this regard, the d$t_{\rm floor, H}$ that primarily operates in 
the inner RSD to limit the Hall diffusivity $\eta_{\rm H}$ 
(left panel of Fig.~\ref{Fig:EtaBc_late}) should not affect qualitatively the 
behavior of the magnetic field, as the field lines inside the RSD become almost 
inert against any tendency of bending (bending can still be efficient near the 
disc surface layers and bipolar regions where $\eta_{\rm O}$ is small). However, 
in studies fully resolving the protoplanetary disc and the magnetic diffusion, 
result may be different quantitatively \citep[e.g.,][]{BaiStone2017}.

Furthermore, the Hall diffusivity $\eta_{\rm H}$ changes its sign to positive 
in the inner RSD, where only electrons are still coupled to the magnetic 
field and dominate the Hall conductivity 
\citep[\S~\ref{Chap.Hall}, see also][]{KunzMouschovias2010}. As a result, 
such a sign change also flips the sign of Hall drift velocities in both 
the radial and azimuthal directions, in comparison to the drift velocities 
in the surrouding envelope (see Fig.~\ref{Fig:sketch-B}). Recall that the radial 
Hall drift $\varv_{{\rm H},r}$ in the RSHCS just outside the RSD is pointing 
inward, the slight azimuthal bending of magnetic fields towards +$\phi$ 
(preferentially following the gas rotation) inside the RSD tends to induce a 
radially outward Hall drift, which may encounter those inwardly drifting field 
lines and produce a pile-up of magnetic flux near the disc edge. However, with 
the Hall d$t$ floor affecting the bulk of the inner RSD, there is only a slight 
indication of such opposite Hall drifts across the disc edge (see bottom panels 
of Fig.~\ref{Fig:2.4opt3-}), clear identification of such a ``Hall flux pile-up'' 
may require simulations that fully resolve the Hall effect inside the RSD. 

%Either model shows a peak of magnetic torque outside their respective 
%disc edges where the poloidal magnetic field $B_z$ is the strongest. 

%The gradual shrink of the inner compartment by self-gravity.
%The inner RSD gradually loses the centrifugal support against gravity and shrinks in 
%radius, while the outer part of the disc flattens vertically as the plasma-$\beta$ 
%there decreases; both are caused by the Hall-induced inward drift of magnetic 
%fields over time.

\subsection{Formation \& Evolution of Counter-rotating Discs with Enhanced Hall Diffusivity: Aligned Configuration}
\label{S.HallDisc+}

\begin{figure*}
\centerline{\includegraphics[width=1.17\textwidth]{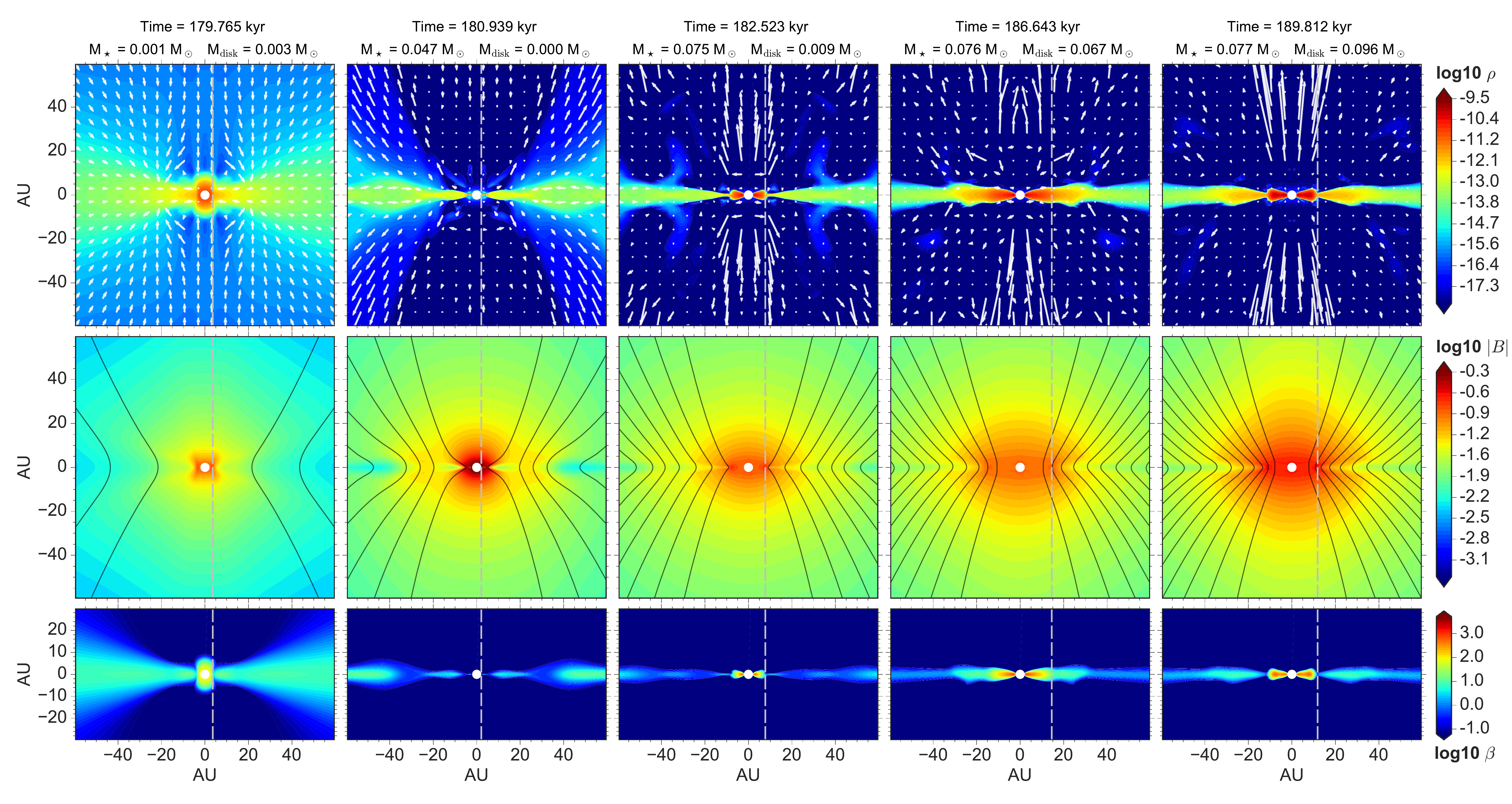}}
\centerline{\includegraphics[width=1.17\textwidth]{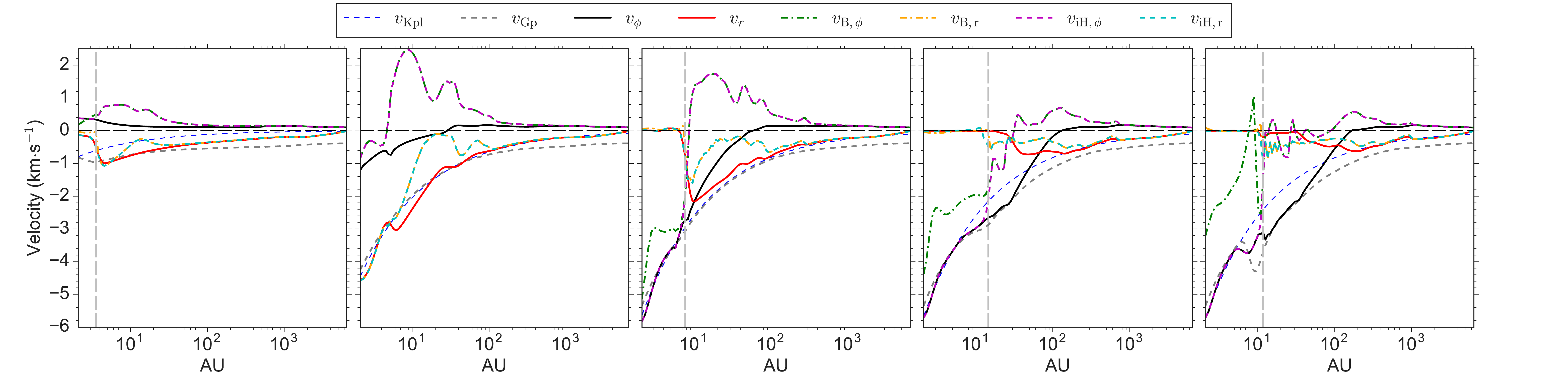}}
\caption{Evolution of disc in the aligned model 2.4opt3-H$^+$O. First row: logarithmic 
distribution of mass density along with velocity field vectors (white arrows). Second row: 
logarithmic distribution of total magnetic field strength $|B|$ along with magnetic 
field lines (black solid lines). Third row: logarithmic distribution of plasma-$\beta$. 
Fourth row: velocity profile along the equator. The vertical silver line (dashed) 
approximately marks the edge of the inner RSD.}
\label{Fig:2.4opt3+}
\end{figure*}

In this section, we focus on the case of aligned configuration ($\bmath{\Omega \cdot B}$>0). 
The suppression of disc formation claimed by previous literature only holds in the 
first $\sim$2~kyr after the first core (Fig.~\ref{Fig:2.4opt3+} for model 2.4opt3-H$^+$O), 
when the enhanced azimuthal bending of magnetic field towards +$\phi$ by the azimuthal 
Hall drift $\varv_{{\rm H},\phi}$ narrowly halts the original gas rotation (along +$\phi$) 
in the inner envelope. As $\varv_{{\rm H},\phi}$ keeps bending the magnetic field 
towards +$\phi$, magnetic tension force towards -$\phi$ increases and large 
counter-rotating motion (``over-shooting'') is able to develop from inside out. 
The strong spin-down torque persists and leads to the formation of a small 
counter-rotating Keplerian disc of a few AU at around $t=182.523$~kyr. Note 
that the only difference between this model and the 2.4MRN-H$^+$O model above 
(Fig.~\ref{Fig:2.4MRN}) is the level of Hall diffusivity; $\eta_{\rm H}$ in this 
model is larger by 1--2 orders of magnitude in the inner 100~AU 
($\sim$10$^{18}$--10$^{19}$~cm$^2$~s$^{-1}$). Hence, the effective azimuthal 
velocity of the magnetic field $\varv_{{\rm B},\phi}$ (or similarly 
$\varv_{{\rm iH},\phi}$) and the gas counter-rotating velocity $\varv_\phi$ 
can reach as high as $\sim$2~km~s$^{-1}$ and -4--6~km~s$^{-1}$, 
respectively, near the central region. 
Up to the point of counter-rotating disc formation, the radial Hall drift of magnetic 
fields induced by the azimuthal bending of magnetic fields (towards +$\phi$) points 
outward along +$r$, i.e., the infall speed of magnetic fields is slower than that of 
the gas at 10--100~AU scale (bottom panels in Fig.~\ref{Fig:2.4opt3+}). Similar to 
the case of anti-aligned configuration, the disc grows in radius in the next 
$\sim$4~kyr due to such a radially outward drift of magnetic fields in addition 
to the efficient spin-down torque generating counter-rotation.

\begin{figure*}
\includegraphics[width=\textwidth]{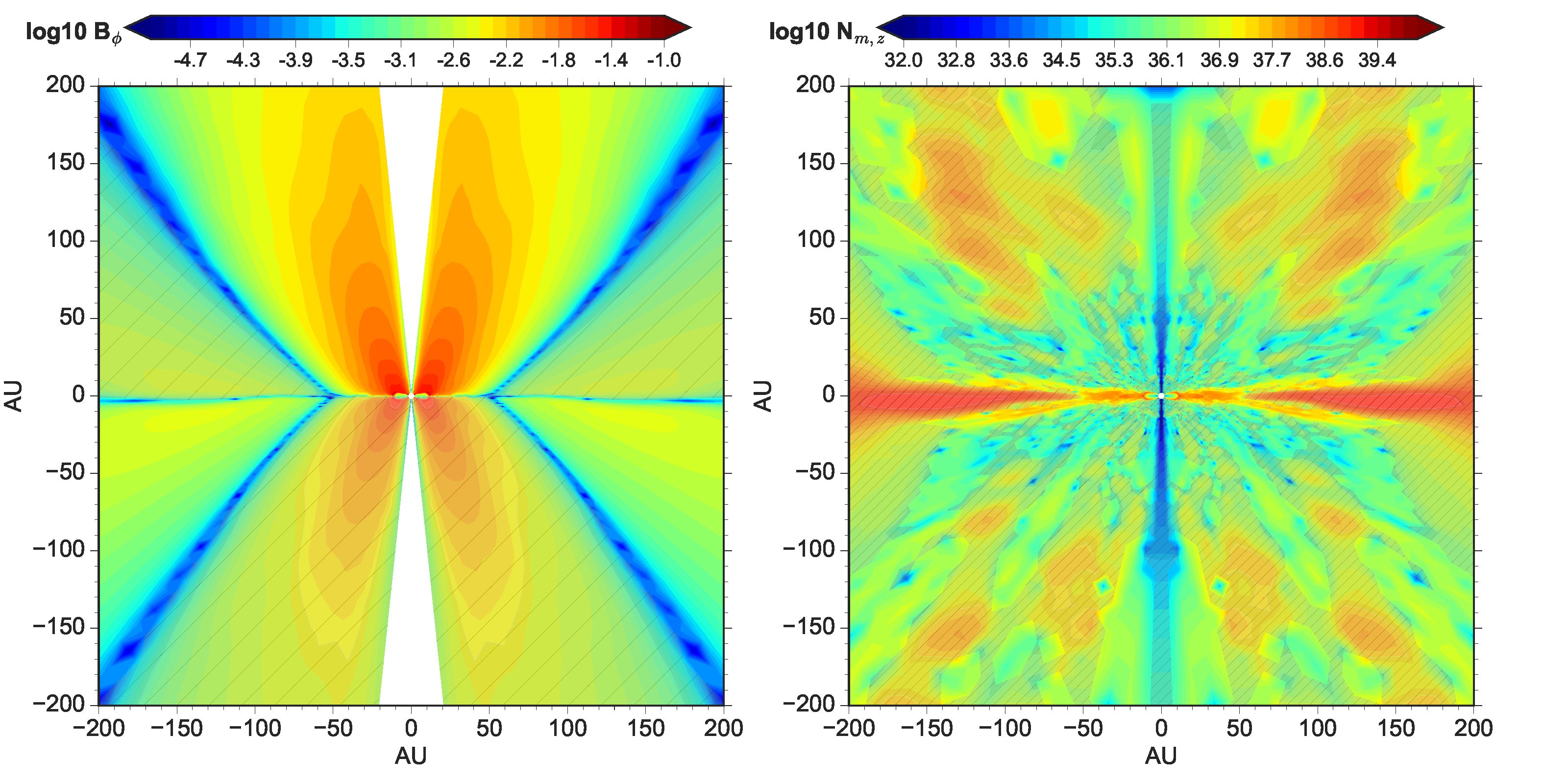}
\caption{Logarithmic distribution of azimuthal magnetic field $B_\phi$ (left panel) 
and magnetic torque $N_{{\rm m},z}$ (right panel) at $t=189.812$~kyr of the aligned 
model 2.4opt3-H$^+$O. $B_\phi$ is positive (along +$\phi$) in the unshaded region and 
negative (along -$\phi$) in the shaded region. Similarly, regions of negative 
magnetic torque (along -$z$) are shown as shaded. Note that the poloidal magnetic 
field points upward in this aligned configuration, and positive torque actually 
denotes torquing down of the counter-rotation.}
\label{Fig:BphTq_align}
\end{figure*}

As the counter-rotating motion dominates the inner tens of AU region, the azimuthal 
Hall drift becomes insufficient to pull the magnetic field along +$\phi$ 
($|\varv_{{\rm H},\phi}|$<$|\varv_\phi|$); instead, the magnetic field starts to 
be dragged azimuthally along the direction of counter-rotation and even to bend 
towards -$\phi$ in the inner tens of AU (Fig.~\ref{Fig:BphTq_align}; see also 
the sketch in Fig.~\ref{Fig:sketch+B}). This will cause the corresponding magnetic 
tension force to operate along +$\phi$, which restrains and torques down the 
counter-rotation motion. Furthermore, the induced radial Hall drift in the inner 
tens of AU also reverses direction to drift the magnetic field radially inward 
with $\varv_{{\rm H},r}\sim-0.5$~km~s$^{-1}$ 
(last two panels in Fig.~\ref{Fig:2.4opt3+}). Similar to the anti-aligned case, 
this outer partition of the disc threaded by highly pinched magnetic fields is a 
RSHCS that flattens vertically over time. After the counter-rotating disc reaches 
a maximum radius around $t=186.643$~kyr, only the inner RSD of $\lesssim$10--15~AU 
radius is long-lived. 

\subsubsection{Counter-rotating Inner Envelope in Aligned Configuration}
\label{S.CounterEnvelope}

\begin{figure}
\includegraphics[width=\columnwidth]{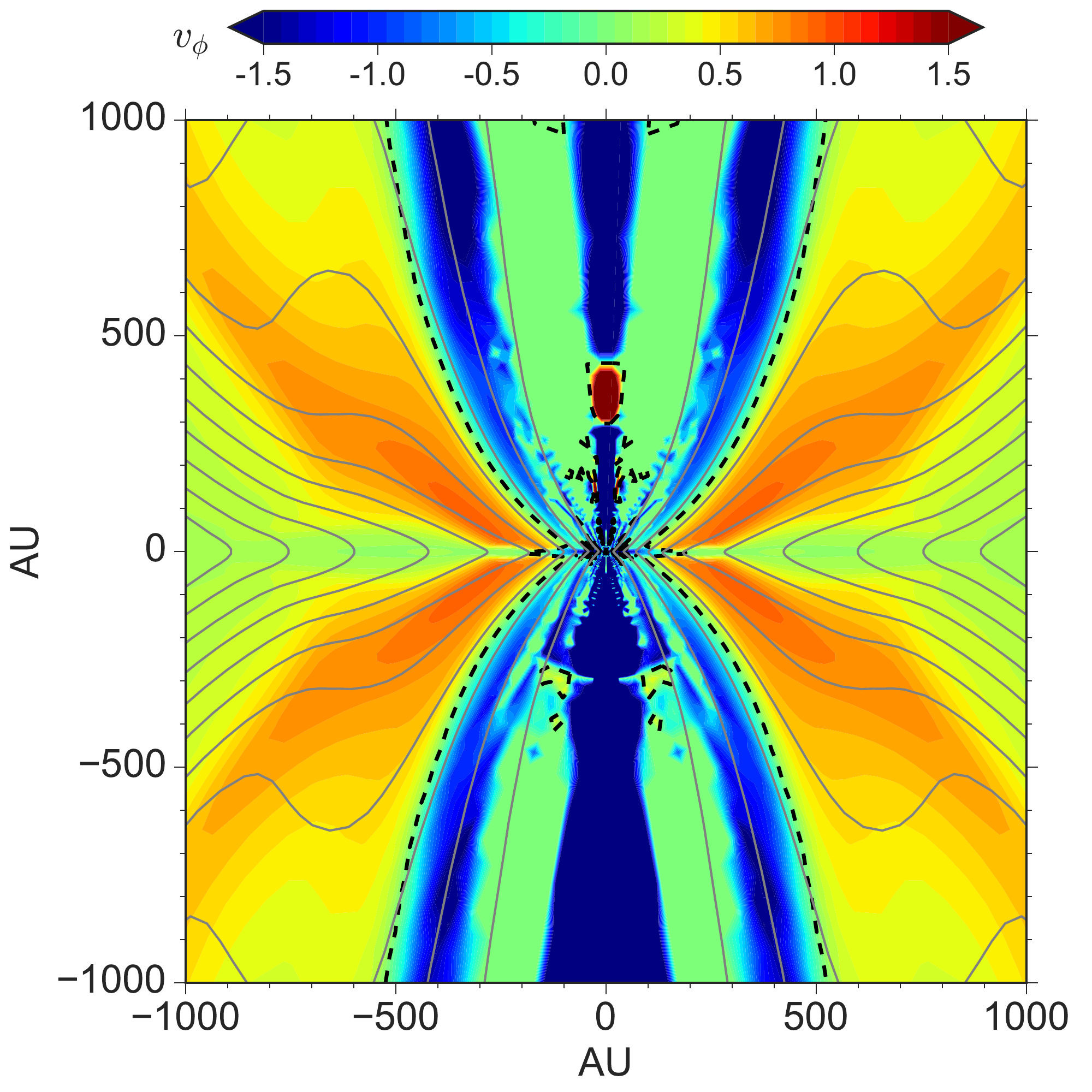}
\caption{Distribution of azimuthal velocity $\varv_\phi$ at 1000~AU scale envelope 
for model 2.4opt3-H$^+$O at $t=189.812$~kyr. Negative $\varv_\phi$ values represent 
rotation motion along -$\phi$ direction. Dashed contour lines mark positions with 
$\varv_\phi=0$, where the transition between positive and negative $\varv_\phi$ 
occurs. Grey solid lines trace the magnetic field lines.}
\label{Fig:Vph_align}
\end{figure}

As the Hall effect is the most efficient in the inner envelope, the region 
with counter-rotating motion is limited to the inner $\sim$200~AU pseudo-disc 
plane (equator in this case) and the associated bipolar outflow cavity 
(Fig.~\ref{Fig:Vph_align}), unlike the thin shell morphology in the anti-aligned 
configuration. The formation of such counter-rotating inner envelope again 
comes from the regulation of the topology of magnetic field lines by Hall effect; 
and naturally, the conservation of angular momentum is not violated. 
The poloidal magnetic field lines at $\sim$1000~AU scale are mostly convex 
(bending towards the equator) in the upper quadrants (or concave in the 
lower quadrants), which induces an azimuthal Hall drift along -$\phi$, 
though small in magnitude, yet accumulatively torquing up the infalling envelope 
within $\lesssim$1000~AU to 1~km~s$^{-1}$ along the original direction of 
rotation (+$\phi$).\footnote{Because the gas rotation $\varv_\phi$ is 
accelerated towards +$\phi$, and the azimuthal Hall drift is relatively small 
$\varv_{{\rm H},\phi}$<0.1~km~s$^{-1}$, the azimuthal bending of B at 
$\sim$1000~AU scale is following the gas rotation along +$\phi$, which further increases 
the curvature of the poloidal magnetic field lines. Therefore, there is no obvious 
transition region of straightening poloidal magnetic curves from convex to concave in 
analogy to the anti-aligned configuration.} 
The excess of angular momentum budget in this part of the envelope balances out the 
negative proportion of angular momentum ($\lesssim$1\% of the total angular momentum 
budget) in the inner counter-rotating part. We again confirm that the total angular 
momentum is mostly conserved in the aligned configuration as well, with a minor 
decrease by $\sim$5\% over a course of $\sim$10~kyr (from 
$1.07\times10^{54}$~g~cm$^2$~s$^{-1}$ at the first core stage to 
$1.01\times10^{54}$~g~cm$^2$~s$^{-1}$ at $\sim$190~kyr for the 
2.4opt3-H$^+$O model) due to the outflow boundary conditions. The picture presented 
here appears to be broadly consistent with the recent observation of counter-rotation 
between disc and envelope around Class I protostar \citep{Takakuwa+2018}. 

\subsection{Effect of Initial Magnetic Field Strength and Rotation Speed}
\label{S.ICimpact}

Across different models we listed in Table~\ref{Tab:model1}--\ref{Tab:model2} 
($\lambda\sim2.4$ and $\lambda\sim4.8$), 
the maximum radius of the disc formed via Hall effect as well as the final radius 
of the inner RSD are relatively insensitive to the initial magnetic field strength 
(for $\lambda\lesssim5$) and rotation speed. Focusing on the models with 
$a_{\rm min}=0.03$~$\mu$m, the final radius of the inner RSD converges to 16--18~AU 
for anti-aligned configuration ($\bmath{\Omega \cdot B}$<0), and 10--12~AU 
(counter-rotating) for aligned configuration ($\bmath{\Omega \cdot B}$>0). 
At intermediate times, the maximum disc radius in the anti-aligned configuration is 
between 40--50~AU with either strong or weak field, and the maximum radius of 
counter-rotating disc in the aligned configuration is between 30--40~AU with strong 
field and 20--30~AU with weak field. Particularly, even when the initial core 
is not rotating, disc can still form by Hall effect with maximum and final radii 
between the aligned and anti-aligned limits. 
This convergence of disc morphology implies that the disc-envelope 
evolution in the main accretion phase is mostly dominated by the self-regulation 
process of Hall effect, which efficiently redistributes angular momentum among 
different parts of the collapsing envelope. As a result, one may derive analytical 
solutions for the equilibrium radius of the inner RSD when including only Hall 
effect and Ohmic dissipation \citep[e.g.,][]{BraidingWardle2012a}. 

\begin{figure}
\centerline{\includegraphics[width=1.0\columnwidth]{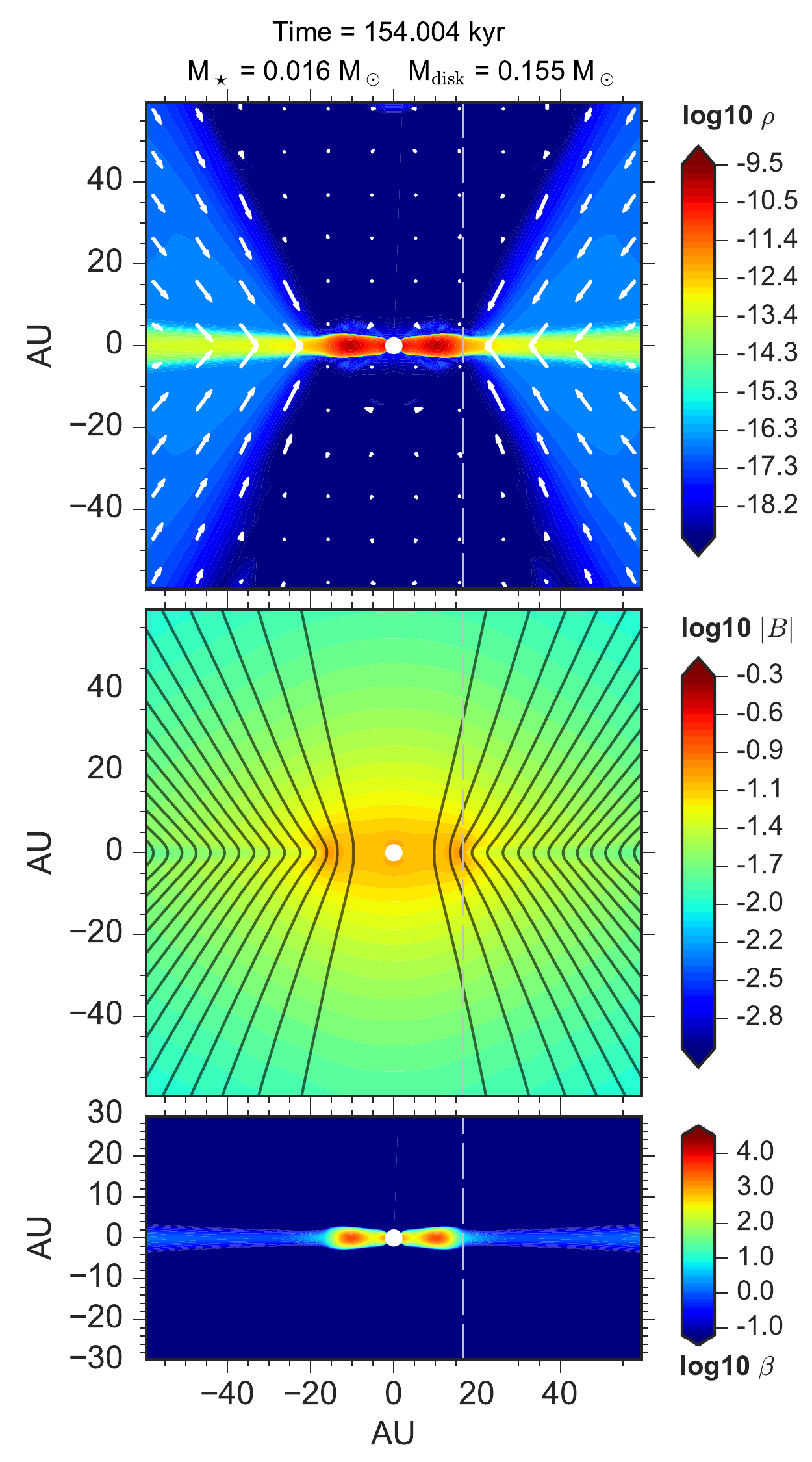}}
\centerline{\includegraphics[width=1.0\columnwidth]{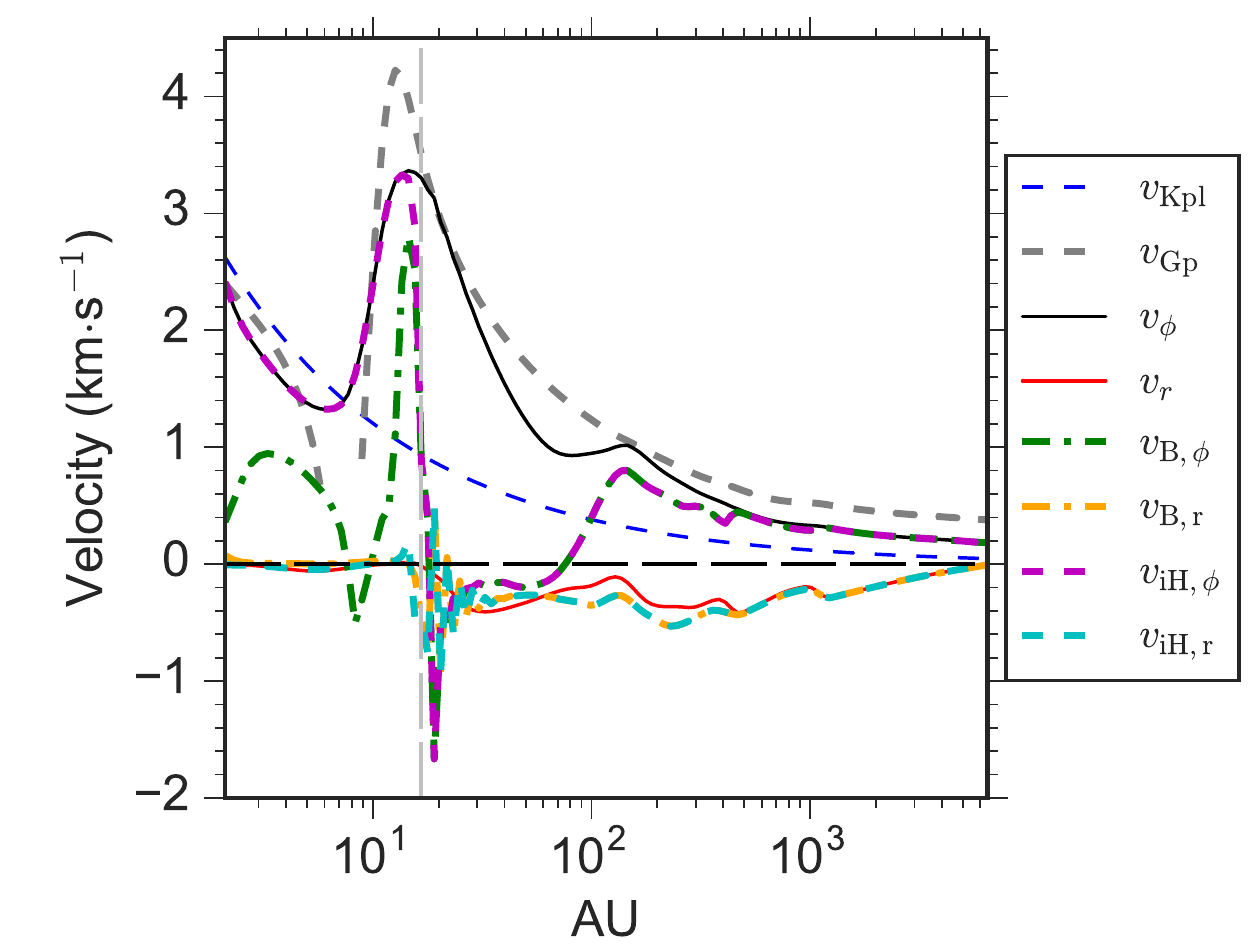}}
\caption{Logarithmic distribution of mass density (first row), magnetic field strength 
(second row) and plasma-$\beta$ (third row), as well as the velocity profile along 
the equator (fourth row) for model 4.8opt3-H$^-$O at $t=154.004$~kyr. White arrows 
in the first row are velocity field vectors and black solid lines in the second row 
are magnetic field lines, respectively.}
\label{Fig:4.8opt3-}
\end{figure}

However, the outer partition of the disc, i.e., the RSHCS (between $\sim$20--40~AU; 
see \S~\ref{S.HallSheet}) is noticeably affected by the magnetic field strength; 
since the RSHCS is intrinsically a pseudo-disc structure, it becomes less prominent 
in the weak field models. In Fig.~\ref{Fig:4.8opt3-}, we present the 4.8opt3-H$^-$O 
model at $t=154.004$~kyr when the total mass of star and disc reaches 0.171~M\sun, 
similar to the frame $t=188.544$~kyr of the strong field model 2.4opt3-H$^+$O we 
discussed above (Fig.~\ref{Fig:2.4opt3-}). By this time, the outer RSHCS exterior 
to the inner RSD ($\lesssim$16~AU) is almost structureless and unrecognizable, 
except that the thin equatorial layer of the RSHCS is still rotationally supported 
with $\varv_\phi\approx\varv_{\rm Gp}$ between 16--30~AU. The plasma-$\beta$ along 
the RSHCS is also below unity; only in the inner RSD the plasma-$\beta$ reaches 
above 10$^3$.
%No over-bending, only azimuthal Hall drift strong. 

Note that, at early times (within a few kyr after the first core), the initial 
conditions do play a role in the disc morphology, especially for the aligned 
configuration ($\bmath{\Omega \cdot B}$>0). With weaker magnetic field and/or 
faster rotation, a larger angular momentum reservoir in the inner envelope is 
able to assemble discs of $\sim$20~AU radius (e.g., model 4.8opt3-H$^+$O and 
4.8min1-H$^+$O) that is rotating along the same direction with the initial cloud 
rotation. However, as the azimuthal Hall drift keeps bending the magnetic field 
towards +$\phi$ and enhancing the spin-down magnetic torque, the initial disc 
shrinks towards the central stellar object over time, and a new counter-rotating 
disc develops and grows from outside ($\gtrsim$5~AU) in the same fashion as 
described in \S~\ref{S.HallDisc+}. Therefore, it is possible that in the aligned 
configuration, the spin direction of the star itself (and materials close to the star) 
can be different from the bulk protostellar disc (and winds/outflows launched from 
the disc) but the same as the outer cloud rotation, provided a weaker magnetic field 
and/or faster rotation initially as well as Hall dominating the non-ideal MHD 
effects. 

Furthermore, we list a few other differences caused by reducing the initial 
cloud magnetization, which generally lowers the efficiency of Hall drift in 
both azimuthal and radial directions.
\begin{enumerate}
\item The azimuthal ``over-bending'' of magnetic fields in the anti-aligned 
configuration is less prominent throughout the simulations of weak field models, 
hence very little outward Hall drift of magnetic fields is present.
\item The counter-rotating shells in the anti-aligned configuration is slightly 
thicker sideways, but tightly surrounds the outflow cavity in the weak 
field models. The maximum counter-rotating speed reached is also somewhat smaller. 
\item The counter-rotating inner envelope in the aligned configuration is spatially 
less extended in comparison to the strong field models.
\end{enumerate}

\begin{figure}
\centerline{\includegraphics[width=1.0\columnwidth]{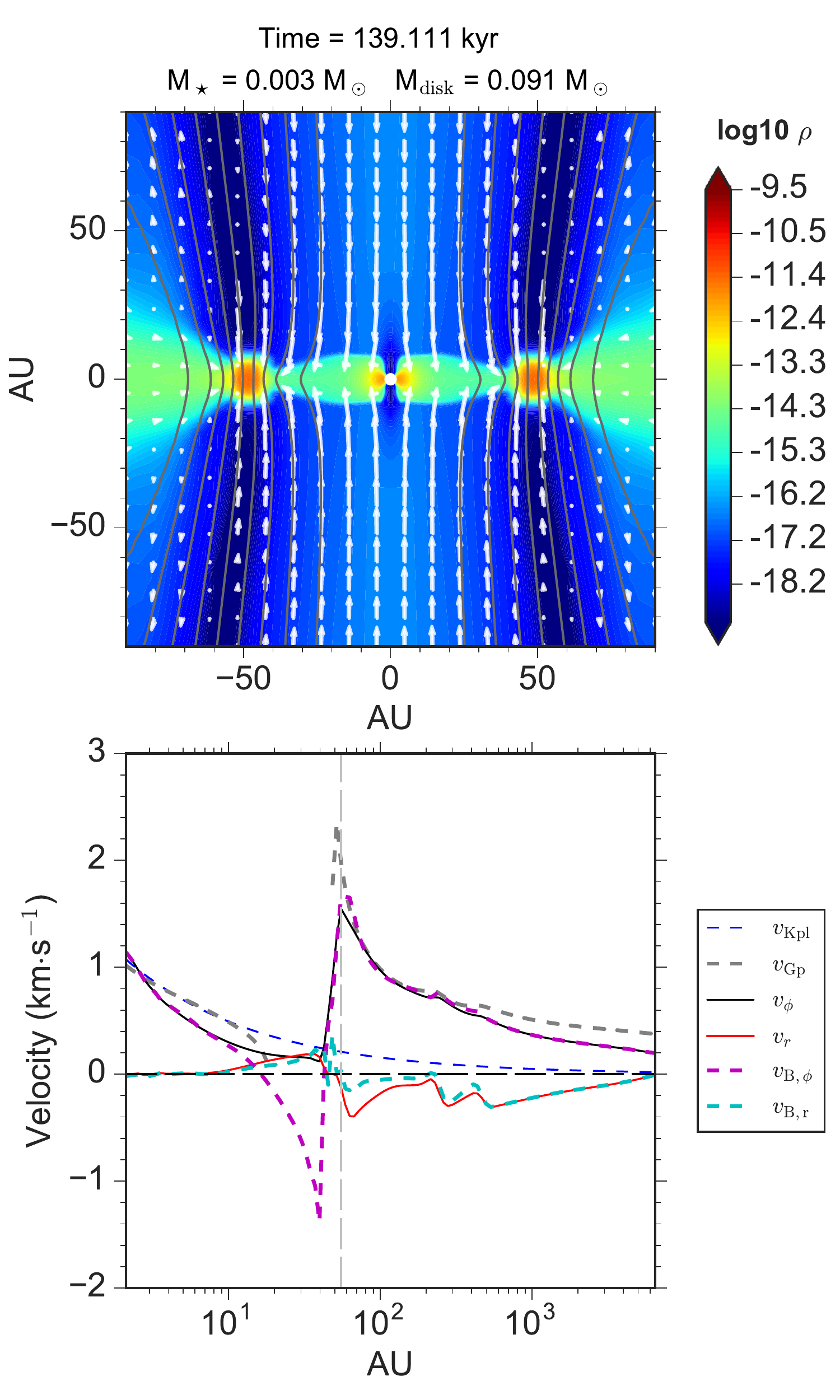}}
\caption{Mass density distribution (top panel) and velocity profile along the 
equator (bottom panel) for model 9.6opt3-H$^+$O. White arrows and grey solid 
lines in the top panel are the velocity field vectors and magnetic field lines, 
respectively. The azimuthal Hall drift is negligible while the radial Hall drift 
diffuses magnetic fields outward.}
\label{Fig:9.6opt3+}
\end{figure}

Finally, for very weak magnetic field $\lambda\sim9.6$ (Table~\ref{Tab:model3}), 
we find the aligned configuration instead produce large ring structures of >50~AU 
that would be a grand design spiral structures in 3D. The poloidal magnetic field is 
only weakly pinched at $\sim$100~AU scale (small ${\partial B_r \over \partial z}$), 
which leads to negligible azimuthal Hall drift of magnetic fields along +$\phi$ 
(Fig.~\ref{Fig:9.6opt3+}). Accordingly, the tension force pointing towards -$\phi$ 
is not much enhanced by such a small $\varv_{{\rm H},\phi}$ to develop obvious 
counter-rotation. On the other hand, the magnetic field is still efficiently pulled 
by rotation towards +$\phi$, which naturally induces a radially outward Hall drift 
reaching 0.3~km~s$^{-1}$ in the inner 100~AU. Such a radial Hall drift behaves 
as an equivalent ambipolar drift that diffuses magnetic flux outward; and the 
disc formation and evolution in model 9.6opt3-H$^+$O is actually similar to the 
AD models we presented in \citet[][see also Appendix.~\ref{App.A}]{Zhao+2016}. 
Therefore, the formation of counter-rotating discs by Hall effect may become 
unlikely for very weakly magnetized cores ($\lambda\gtrsim10$).

\subsection{Lower Limit of Minimum Grain Size \texorpdfstring{$a_{\rm min}$}{} \& 
Upper Limit of Core-Scale Cosmic-Ray Ionization Rate}

As discussed in \S~\ref{S.MRN_LG} above, the Hall diffusivity $\eta_{\rm H}$ 
obtained using the standard MRN size distribution is too small to generate 
efficient Hall drift of magnetic fields; however, as soon as the smallest 
grain population $\lesssim$100~$\AA$ are removed, $\eta_{\rm H}$ can increase 
by >1 order of magnitude in the density range of 10$^9$--10$^{12}$~cm$^{-3}$ 
\citep{Dzyurkevich+2017,Zhao+2018b,Koga+2019}. Such an enhanced $\eta_{\rm H}$ 
is enough to form discs that are comparable to the models with optimal grain 
size $a_{\rm min}$=0.03~$\mu$m. Note that when choosing $a_{\rm min}$=0.01~$\mu$m, 
the values of $\eta_{\rm H}$ at low densities (<10$^9$~cm$^{-3}$) are still 
relatively low \citep[see][]{Zhao+2018b}, but it does not affect the disc 
formation and evolution, because Hall effect only becomes significant within 
$\sim$100--200~AU scale along the equatorial region. Besides, in models 
with $a_{\rm min}$=0.1~$\mu$m, $\eta_{\rm H}$ can briefly become positive at 
an intermediate scale (a few 100~AU) in the envelope 
\citep[see also][]{Dzyurkevich+2017,Zhao+2018b,Koga+2019} but it does not seem 
to affect the morphology and evolution of the disc. Basically, the 
models with $a_{\rm min}$=0.01, 0.03, and 0.1~$\mu$m (corresponds to min1, 
opt3, and trMRN in Table~\ref{Tab:model1}--\ref{Tab:model3}, respectively) 
are very similar in terms of disc size and morphology, indicating that Hall 
effect is less sensitive to grain size distribution than AD \citep{Zhao+2016}, 
as long as the smallest grains ($\lesssim$100~$\AA$) are absent and sub-micron 
sized grains are still present (instead of single sized LGs). In fact, 
recent observations \citep[e.g.,][]{Tibbs+2016} show evidence of depletion of 
nanometer grains ($\lesssim$100~$\AA$) in dense molecular cores. Sticking of VSGs 
onto bigger grains has shown to be rather efficient 
\citep{Ossenkopf1993,Hirashita2012,Kohler+2012}, which is analogous to 
freeze-out of larger molecules on the grain surface \citep{Caselli+1999}.

We have also explored the upper limit of cosmic-ray ionization rate 
$\zeta_0^{\rm H_2}$ (at core scale), above which disc formation by Hall effect 
becomes inefficient. In Table~\ref{Tab:model1}--\ref{Tab:model2}, the models 
with $\zeta_0^{\rm H_2}=10^{-16}$~s$^{-1}$ (10 times the standard value) can 
still form rather compact RSDs of >5~AU radius that is long-lived, in either 
the aligned or anti-aligned configuration. In general, increasing the core-scale 
$\zeta_0^{\rm H_2}$ lowers the overall level of magnetic diffusivities. 
In particular, if $\zeta_0^{\rm H_2} \gtrsim 3 \times 10^{-16}$~s$^{-1}$, 
Hall diffusivity $\eta_{\rm H}$ approaches a level similar to that of MRN models 
with $\zeta_0^{\rm H_2}=10^{-17}$~s$^{-1}$. We therefore test such models with 
$\zeta_0^{\rm H_2} \gtrsim 3 \times 10^{-16}$~s$^{-1}$ (not listed), in which 
small first core type of structures (<5~AU radius) quickly shrink into the 
inner boundary, leaving only pseudo-discs like in the MRN models. Such results 
remain unchanged when adopting different field strengths ($\lambda=4.8$ 
or $\lambda=2.4$) and/or field orientation (aligned or anti-aligned). 
Note that there is little column density increase from the edge of the core 
to the inner envelope, and cosmic-ray attenuation only occurs at disc scale 
given the attenuation length of $\sim$200~g~cm$^{-2}$. Since the most important 
scale for Hall effect to enable disc formation is between a few tens AU to 
$\sim$100--200~AU, hence the Hall diffusivity is still conditioned by 
the un-attenuated $\zeta_0^{\rm H_2}$ at the core scale. Therefore, the upper 
limit of $\zeta_0^{\rm H_2}$ for Hall enabled disc formation should be around 
a few ($\lesssim$3) times 10$^{16}$~s$^{-1}$. However, the upper limit could be 
somewhat increased if asymmetries and perturbations are present in the parent core 
\citep[e.g., misalignment between the initial magnetic field and rotation axis, 
and turbulence close to sonic level; see][]{Joos+2012,Joos+2013,Santos-Lima+2012,Seifried+2013,Li+2013,Li+2014}.

\section{Discussion}
\label{Chap.Discuss}

\subsection{Observational Evidence of Magnetic Field Polarity}
\label{Dis.Observe}

A few recent observations have shown evidence of counter-rotation between disc 
and protostellar envelope \citep[e.g.,][]{Harsono+2014,Takakuwa+2018}. This type 
of sources may better fit with the scenario of aligned configuration between 
the initial magnetic field and angular momentum directions ($\bmath{\Omega \cdot B}$>0). 
The outer envelope is rotating in the same direction as the larger scale cloud, 
while the inner $\sim$200~AU equatorial region and the associated bipolar outflow 
cavity are rotating in the opposite direction (as shown in Fig.~\ref{Fig:Vph_align}). 
The two counter-rotating parts are distinct from each other in both space and in 
kinematics, and thus relatively unambiguous for observational detections. 
On the other hand, the counter-rotating shell in the anti-aligned configuration 
may be more difficult to observe, since the signal of the counter-rotating velocity 
from the thin shell of limited spatial extent is likely to be weak. Furthermore, 
such a velocity component along the line of sight at $\sim$1000~AU scale can 
also be confused with asymmetric infall motions \citep{Yen+2017}.

\subsection{Importance of Magnetic Flux Decoupling From the Collapsing Flow}
\label{Dis.Decouple}

One common characteristic of the discs formed in the Hall dominated cloud cores 
(in the absence of AD) is that, 
despite a brief period of growth to a maximum radius of 40--50~AU after formation, 
their final radius remains small (<20~AU). The outer radii of the disc evolves 
into a flattened pseudo-disc structure, i.e., RSHCS, with large rotation speed induced 
by the azimuthal Hall drift. As we have demonstrated in both the anti-aligned 
(\S~\ref{S.HallDisc-}) and aligned configuration (\S~\ref{S.HallDisc+}), the 
early growth period of the disc coincides with the period that shows prominent 
radially outward Hall drift of magnetic fields. Such an outward drift requires 
the azimuthal bending of magnetic fields to be along the same direction as the 
azimuthal Hall drift (e.g., over-bending of magnetic fields; see 
\S~\ref{S.HallDrift}). As gas rotation dominates the azimuthal Hall drift 
and is able to pull back the bending of magnetic fields azimuthally, the 
induced radial Hall drift then points inward, gradually dragging in more 
magnetic flux and causing the outer part of the disc to evolve into a RSHCS.

Therefore, in order for the RSDs to grow in radius over time, the key still 
lies in a persisting diffusion of magnetic flux radially outward, which is 
unlike to be achieved by Hall effect alone but rather relies on the ambipolar 
drift throughout the collapsing envelope \citep{Zhao+2018a}. We will present 
the combined effect of Hall and AD on disc formation in Paper II.

%Aligned configuration is not as bad as previously claimed, the secondary outward Hall 
%drift is actually helping to regulate the azimuthal enhancement of magnetic braking/tension.

\subsection{Radius of the Inner Boundary}
\label{Dis.Rin}

We also raise caution about our choice of the radius of the inner boundary 
$r_{\rm in}$, which is set to 2~AU, the same as that adopted in \citet{Zhao+2016,Zhao+2018a}. 
As demonstrated in \citet{Machida+2014}, a sink radius of $\lesssim$1--3~AU is ideal 
for resolving the small discs formed in collapse simulations, which can be missed if 
larger sink radii are used (see their Figure 1). However, they also showed that such 
small discs are only transient features that gradually shrink and disappear within 
10$^4$~yrs; but whether a even smaller $\lesssim$1~AU first-core like structure can 
survive and persist is unclear. \citet{Zhao+2018a} confirmed this type of shrinking 
discs and suggested that the phenomena is caused by the large amount of magnetic flux 
dragged into the inner disc forming region, which enhances the magnetic braking of 
the infalling material and reduces the specific angular momentum to be landed on the disc.

In this regard, the choice of $r_{\rm in}$ should not affect much the models forming 
sizable discs of radii larger than $r_{\rm in}$; even at early times, the first cores 
in such models (e.g., 2.4opt3-H$^-$O) are already on the order of $\gtrsim$5~AU. 
However, it is possible that in models of aligned configuration, the transient phase 
with the initial first core shrunk into the inner boundary (e.g., Fig.~\ref{Fig:2.4opt3+}) 
may still keep a tiny disc of $\lesssim$2--3~AU supported by large thermal pressure, 
which could not be resolved in our set-up. The counter-rotating discs then gradually 
grow from outside the normally rotating tiny disc. Similarly, the models showing no 
disc formation (disc radius smaller than $r_{\rm in}$; e.g., the MRN and LG models) 
might instead hold a tiny disc in the very center throughout the Class 0 and Class I 
phase. To fully resolve such tiny disc structures, a proper description of additional 
physical processes may be necessary, including thermal ionization, stellar X-rays, 
and full radiation transfer, etc., which are beyond the scope of this study.

\subsection{Implications of Hall effect in Protoplanetary Disc}
\label{Dis.PPDs}

Although we do not elaborate in this study on the evolution of magnetic fields 
in protoplanetary discs (PPDs) by Hall effect, our result is closely linked to 
such topics. There are already a handful of studies of Hall effect in PPDs with 
both shearing box \citep{Lesur+2014,Bai2014,Bai2015,Simon+2015} and global 
simulations \citep{BaiStone2017}, which show the accretion and wind launching 
strongly depends on the polarity of magnetic fields. 

From our discussion in \S~\ref{S.HallDrift}, it is possible that the vertical 
magnetic fields threading through the PPD may only inherit one polarity 
($\bmath{\Omega \cdot B}$<0) from the collapse of the large scale cloud core, if 
Hall effect already dominates the evolution of the magnetic field in the envelope. 
In Fig.~\ref{Fig:sketch-B}, the anti-aligned configuration results in 
a RSD rotating along the original cloud rotation (+$\phi$) that is 
anti-parallel to the initial magnetic field ($\bmath{\Omega \cdot B}$<0). 
In Fig.~\ref{Fig:sketch+B}, the aligned configuration instead results in 
a counter-rotating disc that is also anti-parallel to the initial magnetic 
field ($\bmath{\Omega \cdot B}$<0). Consequently, the magnetic field along 
the mid-plane of PPDs may preferentially drift outward 
\citep[the second scenario in Fig.~1 of][]{BaiStone2017}, which further 
enhances the rate of magnetic flux diffusion in PPDs.

Furthermore, the negative feedback of Hall drift by its orthogonal Hall drift 
is rarely investigated in the literature of both disc formation and PPD 
evolution. In PPDs, the bending of magnetic fields by Hall drift in the 
radial direction should induce an azimuthal Hall drift that places a 
negative feedback to weaken the original toroidal bending of magnetic fields 
across the disc mid-plane. In \citet{BaiStone2017}, the straightening of 
the poloidal magnetic field lines is attributed to AD, whose effect may 
be of interest to compare with the induced azimuthal Hall drift, especially 
at locations where the radial bending of magnetic fields is severe 
(e.g., disc surface layer). Additionally, Ohmic dissipation can also 
be very efficient in PPD in preventing any tendency of field bending, 
and thus Hall effect may become less important and easily stabilized. 

Nonetheless, all three non-ideal MHD effects are highly dependent on the 
microphysics and ionization chemistry. For example, in the vertical direction, 
the sign of $\eta_{\rm H}$ changes from positive to negative 
(Fig.~\ref{Fig:sketch-B}--\ref{Fig:sketch+B}) across the disc surface layer 
\citep[see also][]{XuBai2016}. 
Hence, the same vertical field line, though bended towards opposite directions 
near the disc mid-plane versus the disc surface layer, may drift together 
towards the same radial direction, %regardless of the disc mid-plane or surface, 
preventing the development of the so-called ``Hall-shear'' instability 
\citep{Kunz2008,BaiStone2017}. Such an effect from the microphysics needs to be 
investigated more carefully in future studies of PPDs.

%The following is naive guess, but should be wrong:
%Both aligned or anti-aligned configuration will torque down the gas flow in 
%disc surface layer and torque up that in the middle plane, which further 
%strengthens the Hall shear. However, in the presence of AD, such instability 
%could be quenched/stabilized \citep{BaiStone2017}. 
%We are not resolving such instabilities in this study, the Hall shear is suppressed 
%in the disc by large Ohmic dissipation; however, our Hall 
%diffusivity is also underestimated by many orders of magnitude inside the disc, 
%hence a small azimuthal bending of magnetic field may induce a large radial Hall drift, 

\subsection{Rapid Magnetic Field Oscillation \& Possible Magnetic Reconnection in RSHCS}
\label{Dis.HallOsc}

Another interesting feature of the RSHCS surrounding the inner RSD is that the 
radial Hall drift $\varv_{{\rm H},r}$ in this region can be fairly oscillatory 
and switches sign rapidly between -$r$ and +$r$. Such a phenomenon is more 
pronounced at later times when a large amount of magnetic flux has been dragged 
into the RSHCS (see Fig.~\ref{Fig:2.4opt3-}, Fig.~\ref{Fig:2.4opt3+} and 
Fig.~\ref{Fig:4.8opt3-}; recall that $\varv_{{\rm H},r}=\varv_{{\rm iH},r}-\varv_r$). 
The radial oscillation of $\varv_{{\rm H},r}$ implies an oscillating direction of 
magnetic field bending azimuthally, which is an outcome of the competition 
between the gas rotation $\varv_\phi$ and the azimuthal Hall drift 
$\varv_{{\rm H},\phi}$. Negative $\varv_{{\rm H},r}$ (pointing inward) 
in the RSHCS increases the degree of radial pinching of magnetic fields and 
amplifies $\varv_{{\rm H},\phi}$ that points to the opposite direction of gas 
rotation in RSHCS. Once such an azimuthal Hall drift is large enough to bend the 
magnetic field in its direction, the induced radial Hall drift $\varv_{{\rm H},r}$ 
becomes positive (points outward). Positive $\varv_{{\rm H},r}$ then reduces the 
radial pinching of magnetic fields as well as the azimuthal Hall drift 
$\varv_{{\rm H},\phi}$, which causes the gas rotation $\varv_\phi$ to dominate 
and to pull back the magnetic field in the direction of rotation, and 
$\varv_{{\rm H},r}$ restores to negative. Generally, gas rotation is large 
enough in the RSHCS to restrain such a brief over-bending of magnetic field 
in the azimuthal direction. 

In essence, the oscillation of magnetic fields in the RSHCS again reveals 
the self-regulating nature of the Hall effect. Once the magnetic fields are 
bended serverely by either the radial or azimuthal Hall drift along the drift direction, 
a secondary Hall drift in the orthogonal direction would be induced,\footnote{\citet{BraidingWardle2012a} 
briefly mentioned such an interdependence in their discussion section.} which 
eventually restrains the original Hall drift. In comparison, relaxing magnetic 
field bending by AD is more straight-forward, in a sense that the direction of 
ambipolar drift is directly opposite to the bending direction of magnetic fields.

Finally, the radial oscillation of Hall drift of magnetic fields in the RSHCS 
can potentially promote magnetic reconnection between poloidal field lines, 
whose foot-points are adjacent to each other along the current sheet (pseudo-disc). 
There is slight evidence of such reconnection mechanisms in our simulations, in 
that magnetic field strength in the RSHCS does not increase drastically with time, 
even with $\varv_{{\rm H},r}<0$ being mostly negative (inward drift) in the RSHCS. 
This type of reconnection may be analogous to the whistler wave mediated fast 
reconnection often discussed in geophysics \citep[e.g.,][]{Mandt+1994,Howes2009}. 
Due to the resistivity floor imposed in the RSHCS, both the oscillation of Hall drift 
and the magnetic reconnection could be underestimated in this study; however, 
including AD (to be discussed in Paper II) will suppress the formation of RSHCS 
along with the rapid oscillation of Hall drift.\footnote{Despite the suppression 
of RSHCS, the main result of the current study still holds in Paper II when all 
three non-ideal MHD effects are considered; i.e., there is no obvious bimodality 
in disc formation.}

\section{Summary}
\label{Chap.Summary}

We have revisited the problem of disc formation by Hall effect, using 2D 
axisymmetric MHD simulations with magnetic diffusivities computed from our 
equilibrium chemical network. We have conducted a parameter study 
(in terms of initial magnetic field strength, rotation speed, cosmic-ray 
ionization rate, as well as grain size distribution), and followed the 
disc-envelope evolution into the main accretion phase. Our main 
conclusions are listed below. 
\begin{description}
\item 1. In collapsing dense cores, Hall effect is inefficient in affecting disc 
formation if the standard MRN size distribution ($a_{\rm min}$=0.005~$\mu$m) 
or singly-sized large grain ($\sim$1~$\mu$m) is adopted for obtaining the Hall 
diffusivity. The Hall drift velocity is too small to affect the azimuthal bending 
of magnetic fields and the resulting magnetic braking. A slightly evolved grain 
size distribution free of VSGs ($\lesssim$100~$\AA$) can greatly boost the Hall 
effect in the inner envelope. 
\item 2. With an optimally enhanced Hall diffusivity corresponding to 
$a_{\rm min}$$\sim$0.03~$\mu$m ($\eta_{\rm H}$ increased by 1--2 order of 
magnitude than that of the standard MRN in the number density range of 
10$^9$--10$^{12}$~cm$^{-3}$), Hall effect can enable the formation of 
compact discs of $\lesssim$10--20~AU radius, regardless of the polarity 
of the magnetic field. 
\item 3. In the anti-aligned configuration ($\bmath{\Omega \cdot B}$<0), 
initial formation of sizable discs of $\sim$30--50~AU radius is enabled by Hall 
effect, with only the inner $\lesssim$10--20~AU being long-lived RSDs and the 
outer region being RSHCSs that flatten over time as the Hall effect 
moves the poloidal magnetic field radially inward relative to matter. 
In the aligned configuration ($\bmath{\Omega \cdot B}$>0), the initial disc 
suppression is followed by the formation of counter-rotating discs 
of $\sim$20--40~AU radius that subsequently evolve similarly to those in the 
anti-aligned case, with the inner $\lesssim$10~AU RSDs being long-lived. 
\item 4. The formation of thin counter-rotating shells at 100-1000~AU scales 
in the anti-aligned configuration is caused by the regulation of magnetic field 
topology by Hall effect. Across the counter-rotating shell, the poloidal magnetic 
curves change from convex- to concave- shaped. In the aligned configuration, 
however, the counter-rotating region is limited to the inner 
$\sim$200~AU pseudo-disc plane and the associated bipolar outflow cavity. 
In either configuration, the angular momentum conservation is not violated; 
Hall effect redistributes angular momentum among different parts of the 
collapsing envelope. %, via regulating the magnetic field topology. 
\item 5. Disc formation via Hall effect in relatively strongly magnetized 
cores ($\lambda \lesssim 5$) is relatively insensitive to the initial magnetic 
field strength and rotation speed, in that the maximum radius of the disc 
and the final radius of the inner RSD converge to roughly similar values 
across different models. However, the outer RSHCS is less prominent in weaker 
field cases due to its pseudo-disc nature. For very weak magnetic field 
$\lambda \gtrsim 10$, the model with aligned configuration evolves similarly 
to an AD model, as the azimuthal Hall drift is negligible while the radial Hall 
drift behaves as an equivalent ambipolar drift that diffuses magnetic flux outward.
\item 6. Besides removing the smallest grains ($\lesssim$100~$\AA$) from the MRN size 
distribution, the cosmic-ray ionization rate at the core scale should not be larger 
than a few ($\lesssim$3) times 10$^{-16}$~s$^{-1}$ in axisymmetric set-ups, in order 
for Hall effect to be efficient in disc formation in collapsing cores.
\item 7. Finally, we point out that the vertical magnetic field threading through 
the PPD may only inherit one polarity ($\bmath{\Omega \cdot B}$<0) from the 
Hall dominated collapsing core, in which case an enhanced rate of magnetic flux 
diffusion is expected in PPDs. We also discuss the rapid oscillation of Hall drift of 
magnetic fields in the RSHCS and possible magnetic reconnection events therein.
\end{description}

We conclude that disc formation enabled by Hall effect is unlikely to be bimodal. 
Either magnetic field polarity can result in compact RSDs of $\lesssim$10--20~AU 
radius; the main difference is the direction of disc rotation. The small disc size 
is closely related to the radially inward Hall drift of magnetic fields over time, 
and is difficult to explain recent observations of a variety of disc structures 
\citep{Perez+2016,Tobin+2016}.
Therefore, to fully solve the puzzles of disc formation, the radially outward 
diffusion of magnetic flux throughout the envelope remains crucial, which is to 
be investigated in details in Paper II. 
%along with Hall effect in Paper II. 

\section*{Acknowledgements}

BZ and PC acknowledge support from the European Research Council 
(ERC; project PALs 320620). 
ZYL is supported in part by NASA 80NSSC18K1095 and NSF AST-1716259 and 1815784. 
KHL acknowledges support from NRAO through an ALMA SOS award. 
HS and RK acknowledge grant support from the ASIAA and the Ministry of Science 
and Technology in Taiwan through MOST 105-2119-M-001-037- and 105-2119-M-001-044-MY3. 
ZeusTW is developed and maintained by the CHARMS group in ASIAA.
Numerical simulations are carried out on the CAS group cluster at MPE.

%%%%%%%%%%%%%%%%% APPENDICES %%%%%%%%%%%%%%%%%%%%%

\appendix

\section{AD+Ohmic Model}
\label{App.A}

\begin{figure}
\centerline{\includegraphics[width=1.0\columnwidth]{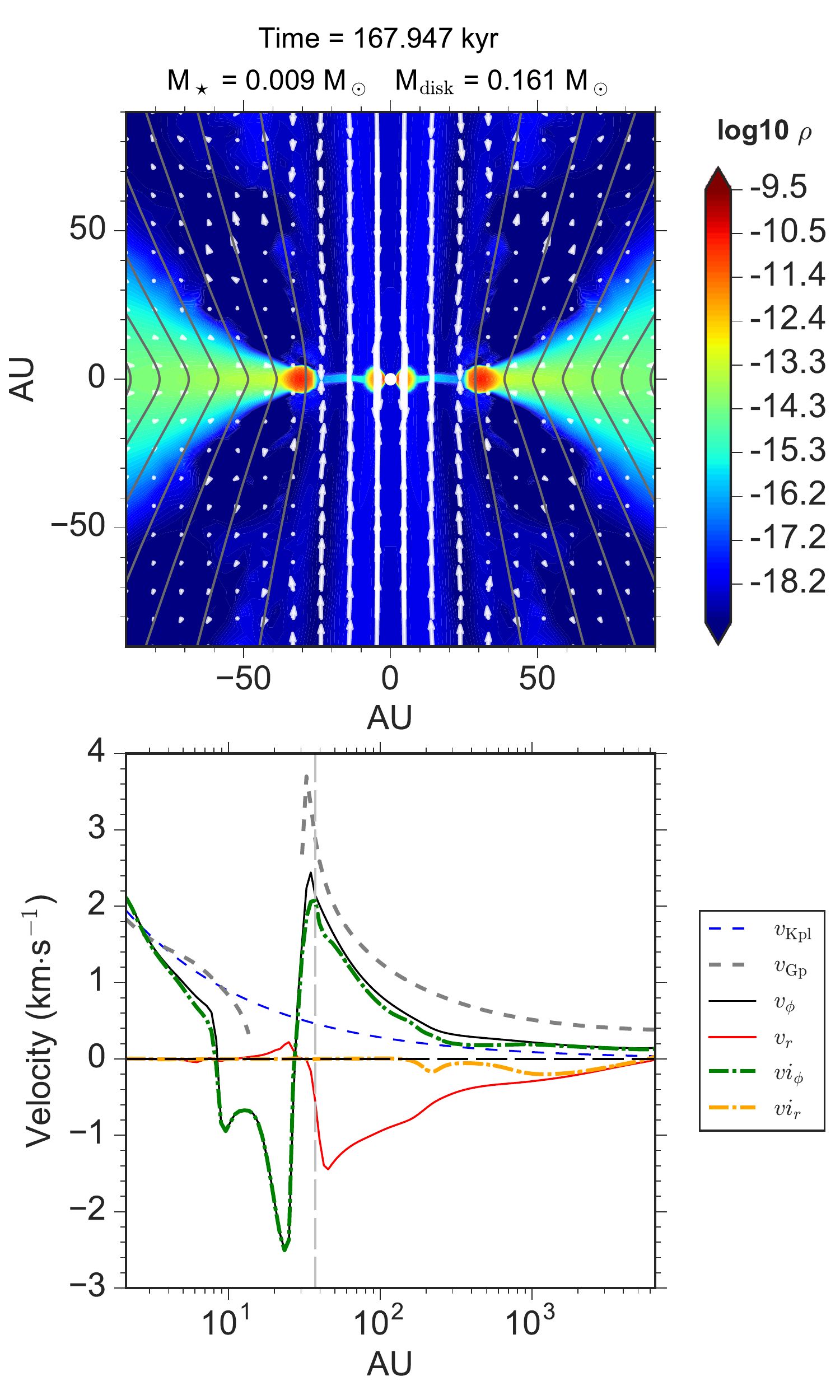}}
\caption{Mass density distribution (top panel) and velocity profile along the 
equator (bottom panel) for model 2.4trMRN-AO with only AD and Ohmic dissipation. 
White arrows and black solid lines in the top panel are the velocity field 
vectors and magnetic field lines, respectively. 
$\varv i_\phi$ and $\varv i_r$ denote the azimuthal and radial velocities of 
the magnetic field lines due to AD (effective ion velocities), respectively.}
\label{Fig:2.4tr-AO}
\end{figure}

In this appendix, we show the AD+Ohmic model 2.4trMRN-AO used to compare with 
the 2.4opt3-H$^-$O model in \S~\ref{S.HallSheet}. The frame at $t=167.947$~kyr 
here shares a similar total mass of star and disc with the frame at $t=188.544$ 
of the 2.4opt3-H$^-$O model. The axisymmetric ring of 30--40~AU 
(Fig.~\ref{Fig:2.4tr-AO}) would actually corresponds to a grand design spiral 
structure in 3D \citep{Zhao+2018a}. 
Although AD induces negligible azimuthal drift of magnetic fields as 
compared to Hall effect, the radially outward ambipolar drift of magnetic fields 
can cause the infall speed of the magnetic field $\varv i_r$ to almost vanish 
within the inner $\lesssim$1000~AU equatorial region, in this 2.4trMRN-AO model 
where $\eta_{\rm AD}$ is optimally enhanced with $a_{\rm min}=0.1$~$\mu$m. 
We refer readers to \citet{Zhao+2018a} for more detailed discussions on the 
radial ambipolar drift.

Note that the criteria for identifying disc in our simulations are similar to 
\citet{Joos+2012} and \citet{Masson+2015}:
\begin{enumerate}
\item Mass density $\rho$ is above certain critical value 
$\rho_{\rm cr} = 3 \times 10^{-13}$~g~cm$^{-3}$; 
\item Azimuthal velocity dominates over radial velocity, i.e., 
$|\varv_\phi|>f_{\rm thres}|\varv_r|$; 
\item The material is close to hydrostatic equilibrium in the $z$-direction, i.e., 
$|\varv_\phi|>f_{\rm thres}|\varv_z|$; 
\item The thermal support dominates over magnetic support, i.e., 
$P_{\rm th}>f_{\rm thres}P_{\rm B}$. 
\end{enumerate}
We choose $f_{\rm thres}=2$ in our analysis throughout the paper.

\section{Constraining Hall \texorpdfstring{\lowercase{d$t$}}{}}
\label{App.B}

\begin{figure}
\includegraphics[width=1.0\columnwidth]{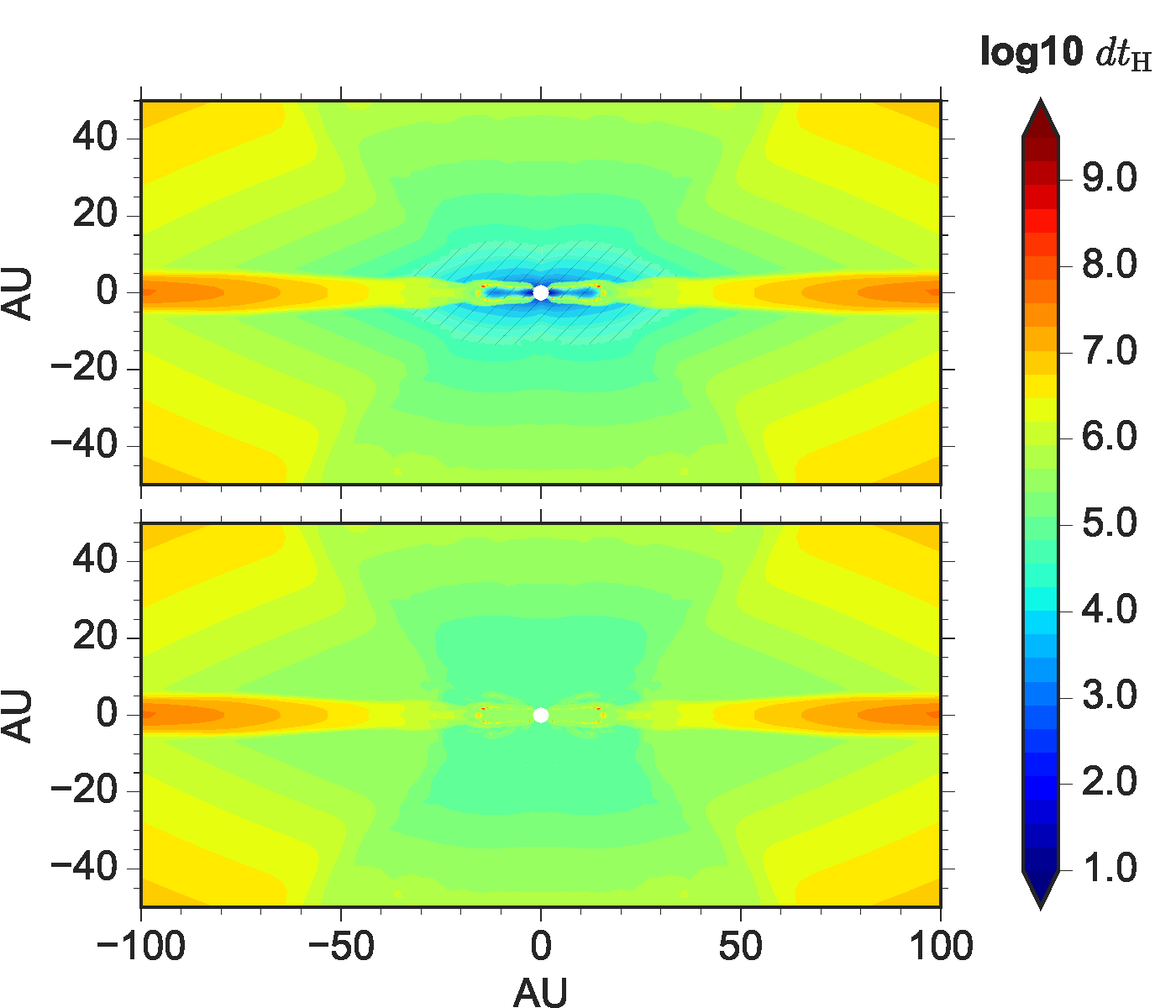}
\caption{Logarithmic distribution of the required Hall time step d$t_{\rm H}$ 
(estimated by $|\delta x|_{\rm min}/c_{\rm w}$) before (top panel) and 
after (bottom panel) applying the $\eta_{\rm H}$ cap (affecting the shaded regions), 
for the model 2.4opt3-H$^-$O at $t=188.544$~kyr. The unit of d$t_{\rm H}$ is in seconds.}
\label{Fig:dtHallfloor}
\end{figure}

In the presence of Hall effect, the Alfv\'{e}n wave splits into left- and right- 
circularly polarized waves with different phase velocities 
\citep[e.g.][]{Wardle1999,SanoStone2002}. The right whistler wave travels much faster 
than the left Alfv\'{e}n wave, with a phase velocity of 
\begin{equation}
c_{\rm w} = {|\eta_{\rm H}|~\pi \over 2 |\delta x|_{\rm min}} + \sqrt{\left({|\eta_{\rm H}|~\pi \over 2 |\delta x|_{\rm min}}\right)^2 + c_{\rm A}^2}
\end{equation}
\citep{Lesur+2014,Marchand+2018}, where $c_{\rm A}$ is the Alfv\'{e}n speed, 
and $|\delta x|_{\rm min}$ is the smallest of the cell's sizes along 
$r$ and $\theta$ directions.
To resolve such a whistler wave using explicit methods, the Hall time step 
$\delta t_{\rm H}$ has to be smaller than $|\delta x|_{\rm min}/c_{\rm w}$, 
which can be a vanishing value and cause the growth of numerical instabilities. 
In this study, we relax such a time step requirement by limiting the Hall diffusivity 
$\eta_{\rm H}$ (see footnote \textsuperscript{\ref{foot:dt_H}}), which essentially 
reduces the whistler wave speed.
As shown in Fig.~\ref{Fig:dtHallfloor}, applying the technique increases the required 
Hall time step d$t_{\rm H}$ from $\sim$10$^2$~s (without $\eta_{\rm H}$ cap) to 
$\sim$10$^4$~s (with $\eta_{\rm H}$ cap) in the inner RSD (dominated by Ohmic) and 
a small region of the bipolar cavity. Along the pseudo-disc and the RSHCS, where most 
of the bending of magnetic field lines and the transport of angular momentum occur, 
d$t_{\rm H}$ is not affected. Therefore, the technique should have little 
effect on disc formation.

%%%%%%%%%%%%%%%%%%%%%%%%%%%%%%%%%%%%%%%%%%%%%%%%%%

% Don't change these lines
\bsp	% typesetting comment
\label{lastpage}
\end{document}